\documentclass[conference]{IEEEtran}
\IEEEoverridecommandlockouts

\usepackage{cite}
\usepackage{amsmath,amssymb,amsfonts}
\usepackage{algorithmic}
\usepackage{graphicx}
\usepackage{textcomp}
\usepackage[svgnames]{xcolor}   
\usepackage{placeins}
\usepackage[hyphens]{url}

\usepackage[table,xcdraw]{xcolor}
\usepackage{tikz}
\usetikzlibrary{arrows.meta,positioning,calc,fit,shapes.geometric,shapes.symbols,intersections}
\usepackage{pgfplots}
\usepgfplotslibrary{fillbetween}
\pgfplotsset{compat=1.18}

\usepackage{float}
\usepackage{enumitem}

\usepackage{siunitx}
\usetikzlibrary{patterns}
\usetikzlibrary{arrows.meta}

\usepackage{comment}
\usepackage{graphicx}
\usepackage{pifont}
\usepackage{diagbox}
\usepackage{subcaption} 

\usepackage{tabularx}
\usepackage{ragged2e}    

\usepackage{pgfplotstable}

\usetikzlibrary{pgfplots.groupplots}
\usetikzlibrary{pgfplots.groupplots,plotmarks}

\usepackage{booktabs}

\pgfplotsset{
  every axis/.append style={
    tick style={black},
    label style={font=\small},
    tick label style={font=\small},
    legend style={font=\small},
    grid style={black!15}
  },
  every axis plot/.append style={
    line width=1pt,
    mark options={scale=0.85}
  }
}

\newcolumntype{P}{>{\hspace{0.3cm}}p{0.9cm}}
\newcolumntype{C}{>{\centering\arraybackslash}m{5cm}}  
\newcolumntype{X}{>{\centering\arraybackslash}m{4cm}}  
\newcolumntype{B}{>{\centering\arraybackslash}p{0.9cm}}  

\def\BibTeX{{\rm B\kern-.05em{\sc i\kern-.025em b}\kern-.08em
    T\kern-.1667em\lower.7ex\hbox{E}\kern-.125emX}}
\begin{document}

\pdfpagewidth=8.5in
\pdfpageheight=11in

\newcommand{\iscasubmissionnumber}{1425}

\newcommand{\scheme}{\textsc{SparseBleed}}
\newcommand{\intronote}[1]{\noindent{\textcolor{red}{[#1] \\}}}
\newcommand{\joshua}[1]{\textcolor{red}{Joshua: #1}}

\pagenumbering{arabic}

\title{\textsc{SparseBleed:} Bypassing Attribute-Inference Defenses via Sparsity Optimizations}

\author{
\IEEEauthorblockN{Darsh Asher\textsuperscript{1*},
Farshad Dizani\textsuperscript{1*},
Joshua Kalyanapau\textsuperscript{1}, \\
Rosario Cammarota\textsuperscript{2},
Aydin Aysu\textsuperscript{1},
and Samira Mirbagher Ajorpaz\textsuperscript{1}}
\IEEEauthorblockA{\textsuperscript{1}\textit{Department of Electrical and Computer Engineering}, \textit{North Carolina State University}, Raleigh, NC, USA \\
\textsuperscript{2}\textit{University of California, Irvine}, Irvine, CA, USA \\
\{dkasher, fdizani, jkalyan, aaysu, smirbag\}@ncsu.edu, rcammaro@uci.edu}
\thanks{\textsuperscript{*}These authors contributed equally to this work.}
}

\maketitle
\thispagestyle{plain}
\pagestyle{plain}

\begin{abstract}
Deep neural networks (DNNs) are widely deployed in safety- and privacy-critical domains such as healthcare and finance, where models are trained on confidential data. This paper presents the first \emph{hardware-level attribute-stealing attack} that bypasses state-of-the-art privacy defenses, revealing that accelerator optimizations intended for performance can directly undermine data confidentiality.



\scheme{} requires no access to gradients, logits, model weights, ground-truth labels, power traces, DVFS manipulation, shared caches, or other channels. Instead, we show that sparsity-driven execution paths encode stable timing signatures that act as a surrogate for model confidence, enabling high-fidelity attribute inference from a single query. \scheme{} generalizes across heterogeneous backends and hardware engines, including Intel AVX2/AVX-512/AMX, NVIDIA Ampere Tensor Cores with 2:4 sparsity, and nine well-known sparse and dense libraries such as Intel oneDNN, oneMKL SparseBLAS, and NVIDIA cuSPARSE.

We evaluate \scheme{} to infer sensitive input attributes and true output labels. On Texas-100X, \scheme{} infers sensitive input attributes such as Sex with 66.39\% accuracy and recovers the true output label, Surgical Procedure, with 99.8\% accuracy.

Finally, we evaluate mitigations. Constant-time execution by disabling sparsity optimizations is expensive (e.g., up to $2\times$ higher latency on NVIDIA Sparse Tensor Cores and $\approx 25\%$ higher power in AMX microbenchmarks). By contrast, a padding defense that equalizes responses to the worst-case time masks timing leakage with only 7.24\% average performance overhead across nine libraries on Texas-100X/Adult and near-zero extra energy.


\end{abstract}

\section{Introduction}
AI models are becoming increasingly sparse \cite{SparseRNNs2023,SparseAttentionMechanism2023,Movementpruning2020,SwitchTransformers2022,SparseGPT2023,SparsitySurvey2018} to improve performance and power consumption. The zero-skipping hardware optimization reduces power consumption \cite{Parashar2017SCNNAA,EIE2016,Cnvlutin}, but the security implications of zero-skipping remain largely unexplored.


As DNNs scale, shared infrastructures and machine-learning-as-a-service (MLaaS) 
deployments have become ubiquitous, relying solely on current software-level mitigations for privacy protection~\cite{Jia2019MemGuard,xue2022use,liu2022opupo,yang2020defending,yang2023purifier}. 
For example, in realistic hospital settings, shared diagnostic models and patient databases are accessed by multiple users and third parties with different privileges, yet all can query the same AI system~\cite{Schneble2020ProjectNightingale, nvidia2023featurestore}. 
These models are trained on proprietary datasets—patient records, financial transactions, or legal documents whose attributes (e.g., diagnoses, procedures, 
or history) are inherently private and must remain protected. 
To preserve privacy, access to inference results or model parameters should not permit reconstruction or identification of such attributes from timing or behavior.



Attribute inference attacks aim to infer sensitive attributes of an individual record that was used to train a model. In the original FJR model-inversion attack (FJRMIA) \cite{Fredrikson2015Confidence}, a black-box adversary supplies the target’s known non-sensitive attributes and systematically varies the candidate values of the sensitive attribute, querying the model for each variant. By combining the confusion matrix with known marginal priors over the sensitive attribute, the adversary infers the most likely true value. Subsequent work \cite{Mehnaz2022Attribute} introduced two variants: a confidence-score-based attack (CSMIA) and a label-only attack (LOMIA). CSMIA uses the model’s confidence scores to infer the sensitive attribute, whereas LOMIA operates in label-only settings with no access to confidence values. LOMIA is similar to FJRMIA but assumes no knowledge of marginal priors over attributes. In the context of medical AI systems, sensitive attributes include medical diagnoses, surgical history, or genetic predispositions; therefore, attribute leakage violates privacy laws like HIPAA and GDPR. 
A common software-level defense against confidence-score-based attribute inference is output discretization—rounding, quantizing, or binarizing the model’s confidence scores to reduce the statistical fingerprint of the training data in the outputs \cite{Shokri2017SP,Jia2019MemGuard,song2021systematic}.

\begin{figure}[!t]
\centering
\scalebox{0.8}{
\includegraphics[width=0.9\columnwidth]{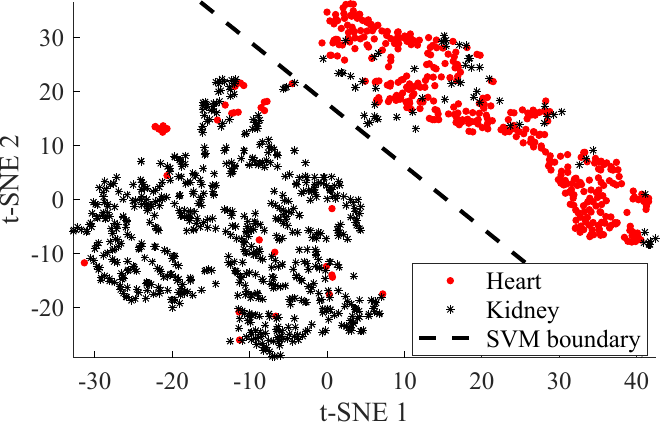}
}
 \caption{Sparsity differences 
 for Heart vs. Kidney in the OrganAMNIST dataset.}
\label{fig:heart_vs_kidney_OrganCMNIST}
\end{figure}

This paper shows, for the first time, that inference time is distinct for different attribute values of user data. Figure~\ref{fig:two_by_one_sparsity_mkl} shows an attribute-dependent difference in induced sparsity within a neural network inference accelerated by Intel SIMD engines (AVX2, AVX-512, AMX). We note further that a higher sparsity results in a lower execution time. We exploit the correlation of sparsity with input attributes in \scheme{} to leak sensitive attributes based on the execution time of neural network inference (Figure~\ref{fig:heart_vs_kidney_OrganCMNIST}). 


This paper evaluates \scheme{} across nine hardware–software combinations, including Intel AMX, Intel AVX‑512, and NVIDIA A100, and libraries such as Intel oneDNN, Intel oneMKL Sparse BLAS, and NVIDIA cuSPARSE.
In all cases, ~\scheme{} successfully leaked sensitive attributes proving the generality of the~\scheme{} side-channel.

Mitigating \scheme{} is challenging because it exploits hardware design, not just software. Disabling zero-skipping results in a power consumption overhead of 25\% and an expensive update that could take years to reach all affected hardware, since there exists no software control for zero-skipping. Output discretization defenses are ineffective since the root of the attack is the inference computation itself, not the model output. 


%

\noindent\textbf{Contributions.} This paper makes the following contributions: 
\begin{itemize}
    \item We introduce \scheme{}, the first hardware-level attribute-inference attack that infers sensitive attributes solely from end-to-end inference time, bypassing probability/score masking and label-only defenses. Unlike timing channels rooted in adaptive early exits \cite{Akinsanya2024TimingCI}, \scheme{} targets non-adaptive networks and does not rely on memory-access pattern monitoring \cite{SparseLeakyNets2024}.


    \item We analyze and quantify how input-dependent activation sparsity, when combined with hardware zero-skipping and software operator/graph optimizations, induces measurable timing differences. We show how this mechanism specifically enables attribute inference in non-adaptive networks and quantify it across nine modern sparse-aware stacks.

    \item We demonstrate that, in sparse-aware libraries/hardware, wall-clock inference time is a reliable proxy for model confidence even without adaptive early exits (contrast with \cite{Akinsanya2024TimingCI}), and we leverage this signal to enable attribute inference under label-only access.


    \item We analyze mitigation strategies and show that full constant-time masking imposes energy 
    overhead ($\approx$25\%)  and would double inference latency whereas padding defense equalizing timing leakage across private attributes gives an average performance overhead of 7.24\% and virtually no energy overhead. 
    
\end{itemize}

\noindent{\textbf{Responsible disclosure. }} We disclosed this vulnerability to Intel, Nvidia, and Arm, who confirmed our findings. 

\noindent{\textbf{Paper Organization. }} Section~\ref{sec:background-real} presents a background on attribute inference attacks. Section~\ref{sec:threatmodel} describes the~\scheme{} threat model. Section~\ref{sec:rootcause} describes the root cause investigation, confirming that the root cause of~\scheme{} is the sparsity acceleration. Section~\ref{sec:sparsebleed-real} describes the~\scheme{} attack. Section~\ref{sec:security-analysis} describes our security analysis of ~\scheme{} on real hardware. Section~\ref{sec:mitigation-real} describes effective mitigations for~\scheme{} and evaluates overhead. Section~\ref{sec:conclusion-real} concludes the paper.

%


\section{Background}
\label{sec:background-real}


\subsection{Attribute Inference Attack}

Attribute Inference attack aims to infer the private sensitive attribute of an individual record from the partial information about the non-sensitive publicly available attributes. By observing the model’s confidence scores \cite{Mehnaz2022Attribute}, the attacker can determine which candidate value of the hidden attribute the model finds most credible. Typical assumptions taken in these types of attacks usually are: (i) The attacker has query access to the model. (ii) The attacker knows the target’s non-sensitive features and true class label. (iii) Optional access to an auxiliary dataset that contains the sensitive attribute\cite{Jayaraman2022Attribute} (or no auxiliary data\cite{Mehnaz2022Attribute}, depending on the variant). 
A query refers to a single interaction with the model—submitting an input and observing the model’s output which mirrors how users typically interact with deployed ML models through APIs.
Here we describe prior attribute-inference attacks.

In Mehnaz et al.'s CSMIA\cite{Mehnaz2022Attribute}, the attacker has black-box query access to the model and can obtain its output confidence scores. For each possible value of the sensitive attribute, the attacker crafts an input consisting of the known non-sensitive attributes augmented with a candidate sensitive attribute value. The attacker queries the model with each variant, records the confidence scores, and selects the attribute value that yields the highest confidence for the true class (i.e., when the prediction matches the known label). Notably, the sensitive attribute decision is made using a simple heuristic and not an additional ML model.

In the shadow-model attribute inference attack\cite{Jayaraman2022Attribute, Mehnaz2022Attribute}, the attacker collects an auxiliary dataset similar to the target model's training set distribution and trains an attack model to predict the sensitive attribute. Attacker knows all the attributes in auxiliary data. If the attacker has access to the target model's confidence outputs, the attacker queries the auxiliary data to collect mappings of the confidence vector to the true attribute . Based on the collected data $(\psi(x),\, \mathrm{shadow\_conf}(x)) \rightarrow t(x)$, attacker trains the attack model to predict the sensitive attribute $t(x)$ where $\psi(x)$ indicates the non-sensitive attributes.  

Otherwise, if the attacker has access only to the output label, the attacker concatenates the output label acquired from the target model to the non-sensitive attributes of the input. The attacker then trains a simple classifier such as logistic regression or a tree-based model on the generated dataset, feeds the victim query into the trained classifier, and uses its prediction as the inferred attribute.

\subsection{Sparsity Implementation and Attacks}
The rising prevalence of sparse models has led to a surge of recently proposed accelerators designed explicitly for efficient sparse computation \cite{bakhtiar2025chason,yadav2025misam,golden2025quartz,gerogiannis2025netsparse,rao2022sparsecore}. Contemporary hardware and runtimes exploit sparsity to meet strict latency and energy budgets \cite{Sze2017EfficientPO,Cnvlutin,EIE2016,Parashar2017SCNNAA,Eyerissv22019,Cambricon-S}. A key mechanism is zero-skipping—the non-execution of operations involving zeros. In practice, zero-skipping is enabled by structured sparsity \cite{StructuredSparsity2016,SparseAttentionMechanism2023}, sparse storage/encodings \cite{EIE2016,Parashar2017SCNNAA}, and on-the-fly zero detection \cite{Cnvlutin}, allowing hardware and libraries to avoid multiply–accumulate operations on zeros; we refer to these collectively as sparsity acceleration. Industry products today like NVIDIA Ampere Tensor Cores with 2:4 structured sparsity \cite{NVIDIAA100,ampereSparsity} with cuSPARSELt support \cite{cusparselt}, and Arm’s Ethos-U microNPUs \cite{armEthosU65} implement sparsity acceleration.

However, sparsity acceleration induces an input-dependent amount of work, which is an attractive property for performance \cite{Sze2017EfficientPO}, but constitutes a potential side-channel. Zero-skipping was shown to be theoretically attackable in the microarchitectural leakage descriptor model proposed in Pandora's Box~\cite{pandorasbox}. 
In SparseLeakyNets \cite{SparseLeakyNets2024}, the authors show that, on sparsity-aware embedded platforms, an attacker with physical or remote access to the device’s external memory bus during inference can infer the model’s predicted label by correlating the timing and volume of transactions with class-dependent sparsity; zero-skipping suppresses memory transactions, yielding class-dependent traffic patterns. In contrast, ~\scheme{}  infers sensitive attributes and true labels using only end-to-end inference time under a black-box, label-only API, with no access to memory traffic or privileged counters.

\scheme{} is fundamentally different from recent AI privacy attacks using hardware optimizations. Unlike Thor~\cite{dizani2025thor}, which leaks the weights of a neural network and works only on single-layer neural networks, ~\scheme{} leaks private attributes of training data to a machine learning model and works on neural networks of any depth. Unlike GateBleed~\cite{kalyanapu2025gatebleed}, which only works on adaptive neural networks~\cite{Akinsanya2024TimingCI,teerapittayanon2016branchynet,skipnet}, ~\scheme{} works on neural networks with a fully static execution path. 
Furthermore, GateBleed is a membership inference attack~\cite{Shokri2017SP,he2025towards,hilprecht2019monte,liu2019performing,gupta2021membership,duddu2020quantifying,liu2021encodermi}. Membership inference attacks leak whether or not an input to an ML model was used in the training set. In contrast, attribute inference attacks leak an attribute of the input to the ML model.

\section{Threat Model}
\label{sec:threatmodel}

Our attack follows the label-only attribute inference setting formalized in prior work~\cite{Fredrikson2015Confidence,Mehnaz2022Attribute}. The adversary targets a standard ML inference API that exposes only the final predicted class label. The adversary is a legitimate user of a black-box inference API, can issue repeated queries, and observes only the final predicted label and the end-to-end latency. The adversary does not have access to model internals (weights, logits, or gradients) and cannot use privileged system side channels (e.g., RAPL or performance counters) or physical channels (e.g., EM or power). The attacker also cannot observe the model’s memory accesses, as assumed by SparseLeakyNets \cite{SparseLeakyNets2024}, and the target model has no adaptive early exits to exploit, unlike \cite{Akinsanya2024TimingCI}. The adversary understands the API schema and the domains of the sensitive attributes and can craft valid queries that fix the target’s known \emph{non-sensitive attributes} while systematically varying the \emph{candidate sensitive attributes} across their possible values. Unlike prior attribute inference settings where the attacker knows the individual’s ground-truth label~\cite{Mehnaz2022Attribute,Kabir2025attribute}, we consider a stronger practical setting in which \emph{both} the sensitive input attribute (e.g., smoker status) and the sensitive output attribute (e.g., procedure type) are unknown. The attacker must infer both purely from (i) non-sensitive attributes, (ii) returned labels, and (iii) end-to-end timing signals. We refer to this as our \textbf{zero-knowledge threat model}, since the attacker begins with no knowledge of either sensitive field. 


We also evaluate sparsity-based inference under two additional targets that each challenge the attacker in different ways, including retrieval-augmented pipelines used in modern production systems such as NVIDIA Merlin~\cite{nvidia2023featurestore} and Microsoft Copilot~\cite{stratton2024introduction}, where backend feature stores may silently inject private attributes that the attacker cannot query or flip. Each scenario challenges the attacker in a unique way. For example, RAG limits the attacker from being able to query crafted values or flip the values for the sensitive unknown attributes. However, we show in Section~\ref{sec:sparsebleed-real} that despite these constraints, exploiting sparsity still enables accurate recovery of sensitive attributes previously assumed to be secure.

\section{Reverse Engineering}
\label{sec:rootcause}
\begin{figure}[h!]
    \centering
        \centering
        \scalebox{0.85}{
        \begin{tikzpicture}
        \begin{axis}[
            width=\linewidth*0.9,
            height=\linewidth*0.35,
            xlabel={Execution cycles},
            ylabel={Density},
            grid=none,
            tick align=inside,
            font=\scriptsize,
            line width=0.9pt,
            legend columns=2,
            legend style={
                at={(0.55,1.05)},
                anchor=south,
                draw=black,
                fill=white,
                font=\footnotesize,
                column sep=6pt
            },
            cycle list name=black white
        ]
    
        \addplot[
            color=red,
            solid,
            line width=0.9pt
        ] table[
            x=x,
            y=y,
            col sep=comma
        ] {Diagrams/proc93_density.csv};
        \addlegendentry{Dermatology}
    
        \addplot[
            color=black,
            dashed,
            line width=0.9pt
        ] table[
            x=x,
            y=y,
            col sep=comma
        ] {Diagrams/proc26_density.csv};
        \addlegendentry{Abdominal}
    
        \end{axis}
        \end{tikzpicture}
        }
    \caption{Inference latency for two different values of a private attribute in Texas-100X dataset.}
    \label{fig:Attribute-inference_relation}
\end{figure}
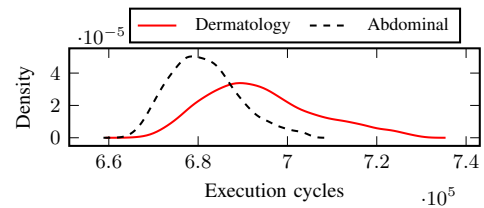

\noindent\textbf{Discovery of attribute-dependent NN inference time.} Our experiments show that the inference time of neural networks can vary significantly and consistently across different sensitive attribute classes.  
Figure \ref{fig:Attribute-inference_relation} shows the inference-time distributions for two procedure-code classes (Abdominal vs. Dermatology) from the Texas-100X dataset, where all non-sensitive attributes are held constant. 
In other words, the sensitive attribute alone changes the effective amount of work executed and shifts total latency by several thousands of cycles.

\subsection{Software-based root-cause examination}
 \noindent{\textbf{Eliminating secret-dependent branch as the root cause.}}
We verified that the network contains no input/secret-dependent branches or dynamic control flow (e.g., Mixture of Experts (MoE) routing or early-exit mechanisms) that could introduce data-dependent timing. The only difference between inputs in Figure \ref{fig:Attribute-inference_relation} is the value of the sensitive attribute.

\begin{figure}[!htbp]
\centering
\scalebox{0.85}{
\begin{tikzpicture}
\begin{axis}[
    width=\columnwidth,
    height=3cm,
    xlabel={Inference Time (CPU cycles)},
    ylabel={Density},
    ymin=0,
    xmin=5.7e5,      
    xmax=6.2e5,      
    ymajorgrids=false,
    grid style={dashed,gray!30},
    tick style={black},
    xlabel style={font=\small},
    ylabel style={font=\small},
    legend columns=2,
    legend style={
    at={(0.5,1.05)},
    anchor=south,
    draw=black,
    fill=white,
    font=\footnotesize,
    column sep=6pt
    },
]

\addplot[
    color=red,
    solid,
    line width=0.9pt
] table[x index=0, y index=1] {Diagrams/curve93.dat};
\addlegendentry{Dermatology}

\addplot[
    color=black,
    dashed,
    line width=0.9pt
] table[x index=0, y index=1] {Diagrams/curve26.dat};
\addlegendentry{Abdominal}

\end{axis}
\end{tikzpicture}
}
\caption{cuSPARSE GPU (A-100): inference timing differences. }
\label{fig:twoclasslarge}
\end{figure}
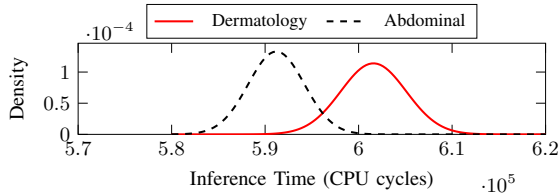

 \noindent{\textbf{Sparsity libraries not required but significantly increase the channel strength.}}
  As Figure~\ref{fig:twoclasslarge} shows, the attribute dependent timing difference becomes substantially
larger once the model is executed through highly–optimized sparse
libraries 
such as oneMKL SparseBLAS and cuSPARSE), the attribute-dependent timing gap increased dramatically 
with
a 16\% increase in mean separation leaving only negligible overlap between the two \textit{timing} distributions of attributes in some cases. 



\noindent\textbf{Eliminating single-architecture behavior.}
Keeping the target attribute fixed, we verify that timing leakage is not tied to any specific model size.
Table \ref{tab:diff_model_leakage} evaluates MLPs spanning depths 2–9 and widths 8–512, and all architectures show measurable timing separation between sensitive attribute classes. Several configurations even yield large effect sizes (e.g., $|d|=0.555$, $0.780$, $0.975$), indicating that the phenomenon persists across model families rather than arising from a single “pathological’’ network.

\begin{table}[!htbp]
\centering
\small
\setlength{\tabcolsep}{4pt}
\renewcommand{\arraystretch}{0.9}
\scalebox{0.85}{
\begin{tabular}{l l S[table-format=6.0] S[table-format=1.3]}
\toprule
\textbf{Width} & \textbf{Depth} & \textbf{Activation} & \textbf{Cohen's d} \\
\midrule
8 & 4 & {288}      & {0.0115} \\
16 & 4 & {1,088}      & {0.0271} \\
32 & 4  & {4,224}     & {0.1045} \\
64 & 4  & {16,640}    & {0.1998} \\
128 & 2   & {33,024}    & {0.0500} \\
128 & {3}   & {49,536}    & {0.2800} \\
{128} & {4} & {66,048} & {0.5552} \\
{128} &{5} & {82,560} & {0.9500} \\
{128}& {6} & {99,072} & {0.7800} \\
{128}&{7}& {115,584} & {0.6500} \\
{256}&{4}& {263,198}   & {0.4767} \\
{512}&{4}& {1,050,624} & {4.7800} \\
\bottomrule
\end{tabular}
}
\caption{Channel strength for various model size. } 
\label{tab:diff_model_leakage}
\end{table}

\begin{figure}[!htbp]
\centering
\scalebox{0.85}{
\begin{tikzpicture}
\begin{axis}[
    width=0.90\columnwidth,
    height=0.35\columnwidth,
    xlabel={Width},
    ylabel={Depth},
    xmin=0, xmax=540,
    ymin=1.5, ymax=7.5,
    grid=none,
    tick align=inside,
    line width=0.9pt,
    enlargelimits=false,
    scatter/use mapped color={
        draw=black,
        fill=mapped color
    },
    scatter,
    only marks,
    mark=*,
    mark size=3pt,
    colormap/viridis,
    colorbar,
    colorbar style={
        ylabel={Cohen's $d$},
        ytick={0,0.5,1},
        yticklabels={0,0.5,1+},
        yticklabel style={font=\scriptsize},
        label style={font=\scriptsize},
    },
    tick label style={font=\scriptsize},
    label style={font=\small}
]

\addplot[
    scatter,
    scatter src=explicit,
] coordinates {
    (  8, 4) [0.0115]
    ( 16, 4) [0.0271]
    ( 32, 4) [0.1045]
    ( 64, 4) [0.1998]
    (128, 2) [0.0500]
    (128, 3) [0.2800]
    (128, 4) [0.5552]
    (128, 5) [0.9500]
    (128, 6) [0.7800]
    (128, 7) [0.6500]
    (256, 4) [0.4767]
    (512, 4) [1.0]
};

\end{axis}
\end{tikzpicture}
}
\caption{Channel strength across architectures: each point represents a (width, depth) feedforward network, colored by Cohen's $d$ for the timing channel.}
\label{fig:width_depth_cohen_heat}
\end{figure}
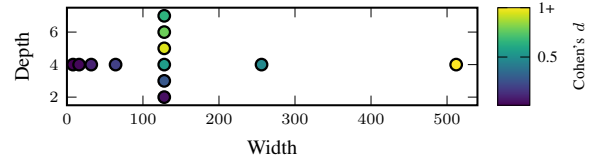

 \noindent\textbf{Impact of model size on the channel strength.} Interestingly, the largest effect size in Table~\ref{tab:diff_model_leakage}
arises not from the deeper 128-wide networks but from the 512-wide
architecture (Cohen’s $d=4.78$). 
This is consistent with the scale effects in NN sparsity: increasing width dramatically increases the absolute
number of \textit{activations} per layer. Figure~\ref{fig:width_depth_cohen_heat} shows Table~\ref{tab:diff_model_leakage} visually - the visibility of the zero-skipping on the tested neural network architectures.  

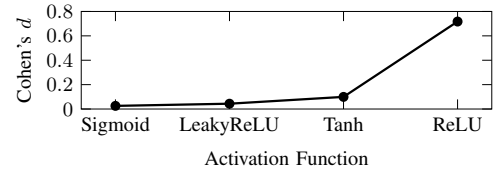
\begin{figure}[!htbp]
\centering
\scalebox{0.85}{
\begin{tikzpicture}
\begin{axis}[
    width=\linewidth*0.9,
    height=\linewidth*0.35,
    ylabel={Cohen's $d$},
    xlabel={Activation Function},
    xtick={1,2,3,4},
    xticklabels={Sigmoid, LeakyReLU, Tanh, ReLU},
    ymin=0.0, ymax=0.8,
    ymajorgrids=false,
    tick style={black},
    legend columns=2,
    legend style={
        at={(0.55,1.05)},
        anchor=south,
        draw=black,
        fill=white,
        font=\footnotesize,                
        column sep=6pt
    },
]

\addplot[
    mark=*,
    thick,
    color=black,
    line width=1pt
] coordinates {
    (1, 0.0257)
    (2, 0.0436)
    (3, 0.0995)
    (4, 0.7173)
};


\end{axis}
\end{tikzpicture}
}
\caption{Cohen's $d$ across activation functions.}
\label{fig:cleaned-cohen-activation}
\end{figure}

\noindent{\textbf{Effect of NN activation function. }}
We compared ReLU with several alternative activation functions that do not induce runtime sparsity. 
ReLU exhibits a substantially higher Cohen’s d, indicating strong distinguishability between sensitive classes, whereas all non-sparsifying activations (e.g., Tanh, Sigmoid, LeakyReLU) yield Cohen’s d values below 0.1, levels that indicate negligible or no separation (see Figure~\ref{fig:cleaned-cohen-activation}). This stark contrast provides direct evidence that sparsity introduced by ReLU is central to the observed timing leakage.

\subsection{Hardware-based root cause examination}
\begin{figure}[!htbp]
    \centering
    \scalebox{0.85}{
    \begin{tikzpicture}
        \begin{axis}[
            font=\scriptsize,
            width=\linewidth,
            height=\linewidth*0.35,
            bar width=50,
            xmin=35500, xmax=56000,
            xlabel={Execution time (cycle)},
            ybar,
            ylabel={Frequency},
            ylabel style={yshift=0pt},
            ymin=0, ymax=0.00205,
            legend style={
                at={(0.5,1.10)}, 
                anchor=north, 
                legend columns=-1,
                font=\scriptsize,
                /tikz/column 2/.style={column sep=5pt}, 
                fill=none,
                draw=none
            },
            legend image code/.code={\draw[fill=#1,draw=none] (0cm,-0.1cm) rectangle (0.25cm,0.1cm);},
            legend entries={100\% sparsity, 50\% sparsity, 0\% sparsity}
        ]
            \addplot+[hist={density, bins=60}, fill=lightgray, draw=none] table[y index=0] {Diagrams/Data/1000-0.txt};
            \addplot+[hist={density, bins=60}, fill=black, draw=none] table[y index=0] {Diagrams/Data/1000-half.txt};
            \addplot+[hist={density, bins=60}, fill=gray, draw=none] table[y index=0] {Diagrams/Data/1000-1.txt};
        \end{axis}
        \node[anchor=south east] at (rel axis cs:-0.05,1) {};
    \end{tikzpicture}
    }
 \caption{Operand-data-dependency in Intel AMX. 
 }
\label{fig:Cumulative_execution_time}
\end{figure}
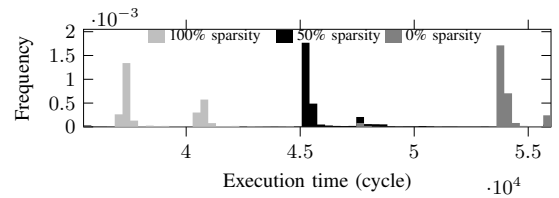

\noindent{\textbf{Reverse engineering operand-value-dependency in Intel AMX.}} 
We execute a test program that performs 1000 AMX multiplications with varying levels of tile operand sparsity (0\%, 50\%, and 100\%). In Intel AMX, a tile is a 2D register that holds a block of data for fast matrix operations. Our results reveal that varying levels of sparsity significantly impact the 
execution time of AMX multiplications, with increased sparsity leading to faster program execution (see Figure \ref{fig:Cumulative_execution_time}).
This discovery uncovers a new value-dependent timing side channel that reveals the sparsity of tile operands in Intel AMX.

\noindent{\textbf{Ruling out HertzBleed Style DVFS effect. }} To determine whether our signal could be explained by Hertzbleed-like frequency scaling, we repeated the experiment while pinning the CPU frequency at 800 MHz, 1.2 GHz, and 1.6 GHz. In each setting, we also measured the energy cost difference between executing an operation on a zero operand versus a maximal (255) operand.

Prior attacks such as Hertzbleed~\cite{Wang2022hertzbleed}, GPU.zip~\cite{wang2024gpu}, Collide+Power~\cite{kogler2023collide+}, and PLATYPUS~\cite{Lipp2021Platypus} exploit operand-dependent instruction (e.g., \texttt{IMUL}) behavior magnified by DVFS to create timing or power leakage. In contrast, our measurements show that such operand-dependent effects have become largely ineffective in recent processors.


\begin{figure}[!htbp]
    \centering
    \scalebox{0.85}{
    \begin{tikzpicture}
        \begin{axis}[
            width=\columnwidth,      
            height=\linewidth*0.35,
            xlabel={\textbf{Frequency}},
            ylabel={\textbf{ 
            Difference (nJ/ops)}},
            xtick=data,
            xticklabels={800MHz, 1.2GHz, 1.6GHz},
            legend columns=2,
            legend style={
                at={(0.5,1.05)},        
                anchor=south,
                draw=black,
                fill=white,
                font=\footnotesize
            },
            tick label style={font=\footnotesize},
            label style={font=\small},
            title style={font=\small, yshift=1.5ex},
            line width=0.9pt,
            mark size=2pt,
            major tick length=0pt,           
            minor tick length=0pt            
            grid=none
        ]

            \addplot[
                color=red,
                mark=*,
                line width=0.9pt
            ] coordinates {
                (1, 0.292057)
                (2, 1.245076)
                (3, 3.011822)
            };
            \addlegendentry{TDPBSSD}
            
            \addplot[
                color=black,
                dashed,
                mark=square*,
                line width=0.9pt
            ] coordinates {
                (1, 0.000000)
                (2, 0.000000)
                (3, 0.000000)
            };
            \addlegendentry{IMUL}

        \end{axis}
    \end{tikzpicture}
    }
    \caption{Mean Energy Difference (nJ/ops) across fixed frequencies for TDPBSSD (AMX FMA operation) and IMUL.}
    \label{fig:energy_diff}
\end{figure}
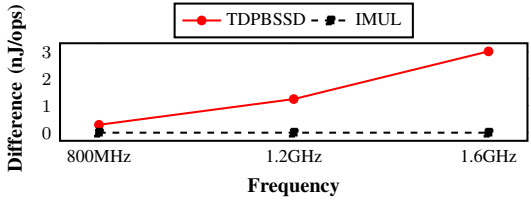
As shown in Figure~\ref{fig:energy_diff},  under a fixed frequency range, there is no operand-sensitive behavior for \texttt{IMUL} operation. 
In contrast, we find that operand-sensitive behavior re-emerges and intensifies within Intel's AMX.
Because the operating frequency is \textit{fixed} in each run, these differences cannot be attributed to data-dependent DVFS. This confirms that the timing and energy variations arise from a mechanism other than Hertzbleed-style frequency changes.

\noindent{\textbf{Reverse Engineering Zero-Skipping in Intel AMX. }}
To test whether AMX employs zero-skipping, we fixed one operand of the \texttt{TMUL} instruction to 127 and swept the other across all 8-bit values (0–255), measuring the energy of 10 million multiplications using RAPL under fixed frequency and thermal conditions.

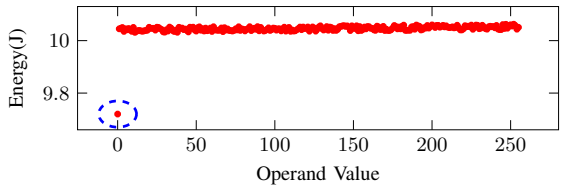
\begin{figure}[!htbp]
    \centering
    \scalebox{0.85}{
    \begin{tikzpicture}
        \begin{axis}[
            xlabel={Operand Value},
            ylabel={Energy(J)},
            only marks,
            width=0.5\textwidth,
            height=3.5cm,
            ymin= 9.66, ymax=10.13
        ]
            \addplot+[red, mark=*, mark options={fill=red}, mark size=1pt] table [x index=0, y index=1, col sep=comma] {Diagrams/Zero_Skipping.csv};

            
            \draw [blue, dashed, very thick]
                (axis cs:0,9.72) ellipse [x radius=12, y radius=0.05];
        \end{axis}
    \end{tikzpicture}
    }
    \caption{Strong evidence of Zero Skipping in Intel AMX}
    \label{fig:operand-energy-sweep}
    \vspace{-0.5em}
\end{figure}

Figure~\ref{fig:operand-energy-sweep} shows a clear pattern: all nonzero operands consume nearly identical energy, while operand $=$ 0 causes a sharp $>$300 mJ drop. This discrete, zero-only reduction cannot be explained by Hamming weight, switching activity, or DVFS, and strongly indicates that AMX skips the multiply when an operand is zero, creating the data-dependent timing channel exploited by our attack.

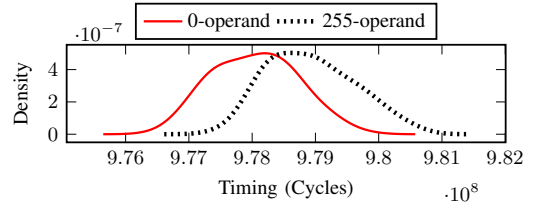
\begin{figure}[!htbp]
    \centering
    \scalebox{0.85}{
    \begin{tikzpicture}
        \begin{axis}[
            width=\linewidth*0.95,
            height=\linewidth*0.35,
            xlabel={Timing (Cycles)},
            ylabel={Density},
            font=\scriptsize,
            grid=none,
            thick,
            legend style={at={(0.5,1.05)},anchor=south,legend columns=-1},
            cycle list={}, 
        ]

        \addplot+[
            red,
            solid,
            line width=1pt
        ]
        table[
            col sep=comma,
            x=x,
            y=pdf,
            restrict expr to domain={\thisrow{type}}{0:0}
        ] {Diagrams/1410_1000_10M_1_1024_ksdensity.csv};
        \addlegendentry{0-operand}

        \addplot+[
            black,
            dotted,
            line width=2pt
        ]
        table[
            col sep=comma,
            x=x,
            y=pdf,
            restrict expr to domain={\thisrow{type}}{255:255}
        ] {Diagrams/1410_1000_10M_1_1024_ksdensity.csv};
        \addlegendentry{255-operand}

        \end{axis}
    \end{tikzpicture}
    }
    \caption{Evidence of Zero-skipping in Nvidia A100 GPU. }
    \label{fig:mma-sync-kernel}
\end{figure}

\noindent{\textbf{Zero-skipping on NVIDIA A100 GPU. }}
Figure~\ref{fig:mma-sync-kernel} exhibits a similar operand-dependent timing difference on a custom CUDA kernel that runs 10 million \texttt{mma\_sync} operations on 1 thread block with 1024 threads. \texttt{mma\_sync} performs matrix multiply–accumulate on small tiles across an entire GPU warp (32 threads), utilizing the A100 Tensor Cores. Notably, the GPU frequency was fixed to 1410 MHz confirming that frequency scaling like Hertzbleed~\cite{Wang2022hertzbleed} is not the root cause. 

\begin{figure*}[!htbp]
\centering
\scalebox{0.85}{
\includegraphics[width=0.9\textwidth]{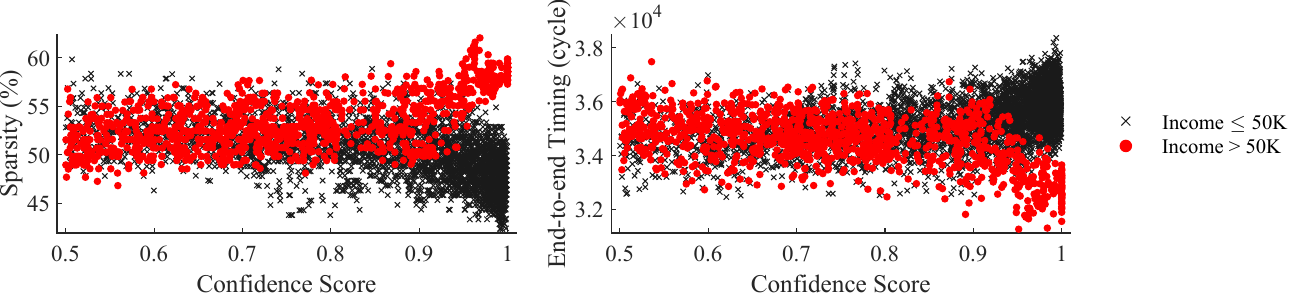}
}
\caption{NN sparsity and inference timing vs. confidence score for different attribute values.}
\label{fig:sparsity_timing_confidence}
\end{figure*}
\noindent{\textbf{Correlation between sparsity, confidence, and timing.}}
We next examine how
attribute-dependent features influence \textit{activation sparsity} and, in turn,
\textit{timing}. In this experiment (Figure~\ref{fig:sparsity_timing_confidence}),
we run the model on the Adult dataset and record three values for every
input: (i) the model’s confidence score, (ii) the activation sparsity
produced inside the network, and (iii) the end-to-end inference time
measured through oneMKL Sparse BLAS.

The left panel of Figure~\ref{fig:sparsity_timing_confidence} shows when confidence is low
($\leq 0.7$), the two sensitive attribute classes (Income $\leq 50$K vs. 
Income $> 50$K) produce very similar internal activations, which in turn
yield nearly identical sparsity levels, leading to overlapping timing
values. As confidence increases, however, the underlying attribute
features induce increasingly different activation patterns, and these
differences propagate into sharply separated sparsity levels. Because
zero-skipping scales directly with the number of non-zero operations,
these sparsity differences are converted into measurable and
attribute-specific timing gaps (right panel).







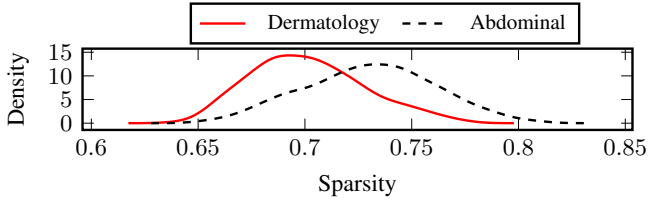
\begin{figure}[htbp!]
\centering






\begin{subfigure}{\linewidth}
    \centering
    \begin{tikzpicture}
    \begin{axis}[
        width=\linewidth,
        height=\linewidth*0.3,
        xlabel={Sparsity},
        ylabel={Density},
        grid=none,
        tick align=inside,
        font=\scriptsize,
        line width=0.9pt,
        legend columns=2,
        legend style={
            at={(0.55,1.05)},
            anchor=south,
            draw=black,
            fill=white,
            font=\footnotesize,                column sep=6pt
        },
        cycle list name=black white
    ]

    \addplot[
        color=red,
        solid,
        line width=0.9pt
    ] table[
        x=x,
        y=y,
        col sep=comma
    ] {Diagrams/proc93_sparsity_density.csv};
    \addlegendentry{Dermatology}

    \addplot[
        color=black,
        dashed,
        line width=0.9pt
    ] table[
        x=x,
        y=y,
        col sep=comma
    ] {Diagrams/proc26_sparsity_density.csv};
    \addlegendentry{Abdominal}

    \end{axis}
    \end{tikzpicture}
\end{subfigure}

\caption{Class-dependent sparsity. }
\label{fig:proc_comparison}
\end{figure}

\noindent{\textbf{Complete causal effect loop from input attribute to timing.}} 
Different sensitive attribute classes induce distinct activation sparsity patterns. As shown in Figure~\ref{fig:proc_comparison}, the Dermatology procedure consistently yields lower sparsity than the Abdominal procedure. Lower sparsity reduces opportunities for zero-skipping, forcing the accelerator to execute more MACs and resulting in higher inference latency (Figure~\ref{fig:Attribute-inference_relation}).
A similar trend appears in Figure~\ref{fig:two_by_one_sparsity_mkl} for Adult-dataset Income classes, demonstrating that attribute-dependent sparsity directly translates into measurable timing gaps across datasets and architectures.

 These results complete the \textit{causal} chain: private attributes induce distinct activations, which drive sparsity, which in turn triggers zero‑skipping and measurable timing differences. \scheme{} leverages this \textit{time-only} signal to infer attributes without confidence outputs.
 
 This \emph{flow} of the leakage
mechanism is: attribute's unique feature $\rightarrow$
activation $\rightarrow$ activation+ReLU $\rightarrow$
sparsity $\rightarrow$ zero-skipping hardware
$\rightarrow$ timing. 

We verified that this attribute–timing correlation persists across the NVIDIA A100 GPU, Intel AMX (on-core matrix accelerator), and x86 vector units (AVX2 and AVX-512).

\begin{figure}[!htbp]
\centering

\pgfplotstableread{Diagrams/Data/Sparsity_y_1.txt}\Syone
\pgfplotstableread{Diagrams/Data/Sparsity_x_1.txt}\Sxone
\pgfplotstablecreatecol[
copy column from table={\Sxone}{[index] 0}
]{x}\Syone

\pgfplotstableread{Diagrams/Data/Sparsity_y_2.txt}\Sytwo
\pgfplotstableread{Diagrams/Data/Sparsity_x_2.txt}\Sxtwo
\pgfplotstablecreatecol[
copy column from table={\Sxtwo}{[index] 0}
]{x}\Sytwo

\pgfplotstableread{Diagrams/Data/Timing_InteloneMKL_y_1.txt}\IMKLyone
\pgfplotstableread{Diagrams/Data/Timing_InteloneMKL_x_1.txt}\IMKLxone
\pgfplotstablecreatecol[
copy column from table={\IMKLxone}{[index] 0}
]{x}\IMKLyone

\pgfplotstableread{Diagrams/Data/Timing_InteloneMKL_y_2.txt}\IMKLytwo
\pgfplotstableread{Diagrams/Data/Timing_InteloneMKL_x_2.txt}\IMKLxtwo
\pgfplotstablecreatecol[
copy column from table={\IMKLxtwo}{[index] 0}
]{x}\IMKLytwo

\begin{tikzpicture}
\begin{groupplot}[
group style={group size=1 by 2, vertical sep=40pt},
width=.95\columnwidth,
height=.3\columnwidth,
grid=none,
ylabel style={font=\small},
label style={font=\small},
tick label style={font=\scriptsize},
legend style={font=\scriptsize, legend columns=2, at={(0.5,1.05)}, anchor=south},
cycle list name=color list,
every axis plot/.style={line width=1pt, mark=none, solid},
unbounded coords=discard,
]

\nextgroupplot[
xlabel={Sparsity (\%)},
ylabel={Frequency},
]
\addplot+[mark=none, color=red!60] table[x=x, y index=0] {\Syone};
\addlegendentry{Income $\leq$ 50K}
\addplot+[mark=none, color=black] table[x=x, y index=0] {\Sytwo};
\addlegendentry{Income $>$ 50K}

\nextgroupplot[
xlabel={End-to-end Inference Time (cycle)},
ylabel={Frequency},
]
\addplot+[mark=none, color=red!60] table[x=x, y index=0] {\IMKLyone};
\addplot+[mark=none, color=black] table[x=x, y index=0] {\IMKLytwo};

\end{groupplot}
\end{tikzpicture}

\caption{Distributions  of sparsity (top)  and timing (bottom). }
\label{fig:two_by_one_sparsity_mkl}
\end{figure}
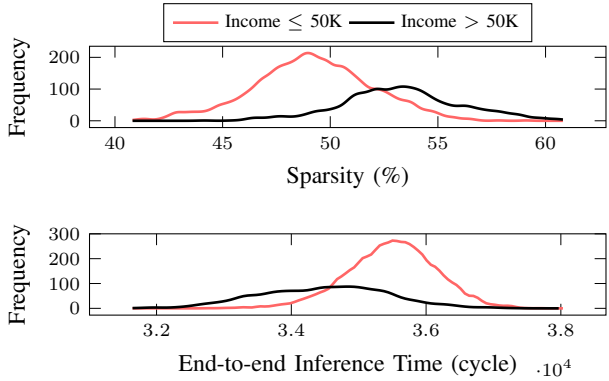

\begin{figure*}[!t]
\centering
\scalebox{0.88}{
\begin{tikzpicture}[
  font=\footnotesize,
  node distance=10mm and 12mm,
  line/.style={-Latex, thick},
  block/.style={draw, rounded corners, align=center, inner sep=6pt, fill=gray!15},
  dataset/.style={draw, cylinder, shape border rotate=90, aspect=0.25,
                   minimum height=16mm, minimum width=35mm,
                   align=center, inner sep=4pt, fill=gray!5},
  api/.style={draw, thick, rounded corners, align=center, inner sep=8pt, fill=gray!45},
  data/.style={draw, dashed, align=center, inner sep=6pt, fill=gray!0},
  decision/.style={draw, diamond, aspect=2, align=center, inner sep=2pt, fill=gray!35},
  phase/.style={draw, dashed, rounded corners, inner sep=8pt}
]

\node[dataset, minimum width=25] (DA) at (-5.2,7.6) {$\mathcal{D}_{A}=\{(\mathbf{x}^{(j)}_{\mathrm{ns}}, s^{(j)})\}_{j=1}^{N}$\\[4pt]
Auxiliary inputs};

\node[api, minimum width=15mm] (API1) at (.1,7.6) {Label-only API\\[5pt]
$\hat{y}=f(\mathbf{x})$\\[3pt] $\ \bar{\tau}=T(\mathbf{x})$};

\node[dataset, minimum width=25] (DT) at (4.7,7.6) {$\mathcal{D}_{T}=\{(\bar{\tau}^{(j)},\hat{y}^{(j)})\}_{j=1}^{N}$\\[4pt]
Timing dataset};

\node[block, minimum width=25] (train) at (8.9,7.6) {Train $g_{\theta}:\mathbb{R}\rightarrow\Delta(|\mathcal{Y}|)$\\[1pt]
$g_{\theta}(\bar{\tau})[c]\approx \mathbb{P}(\hat{y}=c\mid\bar{\tau})$};

\draw[line] (DA) -- node[above] {$\mathbf{x}^{(j)}=(\mathbf{x}^{(j)}_{\mathrm{ns}}, s^{(j)})$} (API1);
\draw[line] (API1) -- node[midway, above, align=center]
{measure $\bar{\tau}^{(j)}$\\[3pt]observe $\hat{y}^{(j)}$} (DT);
\draw[line] (DT) -- (train);

\node[phase, fit=(DA)(API1)(DT)(train)] (phaseI) {};
\node[anchor=west] at ([xshift=1mm,yshift=3mm]phaseI.north west) {\bf Phase I: Learn timing meta-model $g_{\theta}$};

\node[block] (victim) at (-5.3,4.6) {\textbf{Victim}: known $\mathbf{x}^{*}_{\mathrm{ns}}$\\Sweep $s\in\mathcal{S}$\\
Form $\mathbf{x}^{*}_{s_i}=(\mathbf{x}^{*}_{\mathrm{ns}}, s_i)$\\
$\quad i=1,\dots,k$};

\node[data, right=of victim] (sweep) {
$\mathbf{x}^{*}_{s_1}=(\mathbf{x}^{*}_{\mathrm{ns}}, s_1)$\\
$\mathbf{x}^{*}_{s_2}=(\mathbf{x}^{*}_{\mathrm{ns}}, s_2)$\\
...\\
$\mathbf{x}^{*}_{s_k}=(\mathbf{x}^{*}_{\mathrm{ns}}, s_k)$};

\node[api, minimum width=15mm] (API2) at (2.5,4.6) {Label-only API\\[5pt]
$\hat{y}_{s}=f(\mathbf{x}^{*}_{s})$\\[3pt] $\ \bar{\tau}_{s}=T(\mathbf{x}^{*}_{s})$};

\node[data, right=of API2, xshift=14mm] (aggregate) {$ (\bar{\tau}_{s_1}, \hat{y}_{s_1}) $\\
$ (\bar{\tau}_{s_2}, \hat{y}_{s_2}) $\\
...\\
$ (\bar{\tau}_{s_k}, \hat{y}_{s_k}) $};

\draw[line] (victim) -- (sweep);
\draw[line] (sweep) -- node[above] {$\mathbf{x}^{*}_{s_i}$} (API2);
\draw[line] (API2) -- node[midway, above, align=center]
{measure $\bar{\tau}_{s_i}$\\[3pt]observe $\hat{y}_{s_i}$} (aggregate);

\node[phase, fit=(victim)(sweep)(API2)(aggregate)] (phaseII) {};
\node[anchor=west] at ([xshift=1mm,yshift=3mm]phaseII.north west) {\bf Phase II: Victim-specific probing (data collection)};

\node[block] at (8.9,1.8) (score) {$\displaystyle \pi_{s}=g_{\theta}(\bar{\tau}_{s})[\hat{y}_{s}]$};

\node[data, left=18mm of score] (scorevalues) {$ \pi_{s_1} $\\
$ \pi_{s_2} $\\
...\\
$ \pi_{s_k} $};

\node[decision, left=18mm of scorevalues] (argmax) {$\arg\max_{s\in\mathcal{S}} \pi_{s}$};

\node[block, left=12mm of argmax] (output) {Output\\
$\hat{s}^{*}$ and $\hat{y}^{*}=\hat{y}_{\hat{s}^{*}}$\\
(confidence $=\pi_{\hat{s}^{*}}$)};

\draw[line] (aggregate) -- (score);
\draw[line] (score) -- (scorevalues);
\draw[line] (scorevalues) -- (argmax);
\draw[line] (argmax) -- (output);

\node[phase, fit=(score)(argmax)(output)] (phaseIII) {};
\node[anchor=west] at ([xshift=1mm,yshift=3mm]phaseIII.north west) {\bf Phase III: Joint inference (decision rule)};

\draw[line, dashed] (train.south) |- ++(0,-2.0) -| node[pos=0.6,right] {$g_{\theta}$} (score.north);

\end{tikzpicture}
}
\caption{Attack overview with separated phases. Phase~I: construct auxiliary inputs $\mathcal{D}_{A}$ and query the label-only API to obtain timing–label pairs $(\bar{\tau}^{(j)},\hat{y}^{(j)})$ forming $\mathcal{D}_{T}$, then train the timing meta-model $g_{\theta}(\bar{\tau})$ that maps latency to a label distribution. Phase~II: fix the victim’s non-sensitive attributes $\mathbf{x}^{*}_{\mathrm{ns}}$, enumerate $k$ candidates $s_{1},\dots,s_{k}$, form $\mathbf{x}^{*}_{s_i}=(\mathbf{x}^{*}_{\mathrm{ns}}, s_i)$, and collect $(\bar{\tau}_{s_i},\hat{y}_{s_i})$ for each $i$. Phase~III: compute $\pi_{s_i}=g_{\theta}(\bar{\tau}_{s_i})[\hat{y}_{s_i}]$ for all $i$, select $\hat{s}^{*}=\arg\max_{s\in\mathcal{S}}\pi_{s}$, and output $\hat{y}^{*}=\hat{y}_{\hat{s}^{*}}$ with confidence $\pi_{\hat{s}^{*}}$. Here $\bar{\tau}$ denotes the end-to-end latency (mean over repeats if averaging is used).}

\label{fig:attack-overview}
\end{figure*}
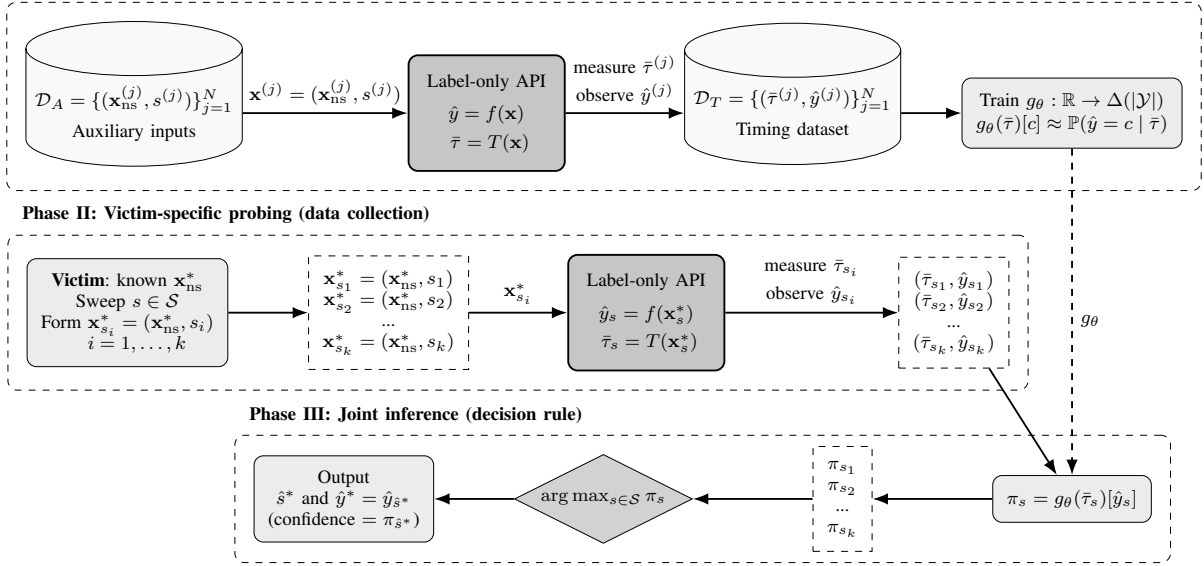

\section{\scheme{}}
\label{sec:sparsebleed-real}

Having identified the root cause, we now shift to describing the technique to use this input-dependent latency from the hardware timing channel to leak sensitive attributes under strict limitations which prior methods fail.


Based on Figure~\ref{fig:sparsity_timing_confidence}, activation
sparsity behaves as an effective proxy for model \textit{confidence}. This
creates a fundamental vulnerability: even if software defenses such as
confidence masking are deployed, an adversary can bypass them by
inferring confidence \emph{indirectly}. Sparsity reveals confidence, and
hardware zero-skipping makes sparsity observable through timing. As a
result, masking the confidence output does not eliminate the underlying
leakage pathway.

Our reverse-engineering results showed that (1) data features correlate
with sparsity due to representation learning in NNs, (2) activation
sparsity is tightly coupled with model confidence, and (3) sparsity
correlates with execution time due to zero-skipping and related
optimizations. Therefore, variations in runtime can inadvertently leak
information about the input data through side-channel timing analysis,
revealing sensitive attributes of the data. The timing histograms were
stable across repeated experiments, and Table~\ref{tab:diff_model_leakage}
quantified the separation using Cohen’s $d$.

We now present three attacker targets as concrete scenarios for \scheme{} that exploit the same root cause:

\begin{enumerate}
\item \textbf{Target A:} inferring \emph{sensitive input attribute} (e.g., Sex, Ethnicity, Source of Admission).
\item \textbf{Target B:} inferring the \emph{true output label} (e.g., procedure code, disease class).
\item \textbf{Target C:} inferring attribute in \emph{retrieval-based systems}, where a backend store retrieves and injects records (e.g., images or tables) that introduce private attributes at inference time.
\end{enumerate}

Each scenario follows the same causal chain but differs in the attacker’s target and how the timing information is used.

\subsection{Target  A: Inferring Sensitive Input Attribute}
\label{sec:attack-input}

\paragraph*{Goal}
Infer a sensitive \emph{input} attribute of a specific individual record (e.g., Sex or Source of Admission in Texas-100X, Marital Status or Race in Adult) by observing the \emph{timing} of the queries.

\paragraph*{Assumption} Compared with CSMIA \cite{Mehnaz2022Attribute}, we assume no access to confidence scores. In contrast to FJRMIA \cite{Fredrikson2015Confidence} and LOMIA \cite{Mehnaz2022Attribute}, we also assume no knowledge of the confusion matrix or marginal priors. Finally, unlike prior attacks, the attacker does not know the ground-truth labels for the attack dataset. Since prior label-only attacks require ground-truth labels to infer the sensitive attribute, they fail in our setting.

\paragraph*{Phase I: Learn a timing model $g_{\theta}$ (latency → label)}
The attacker constructs auxiliary queries that follow the API schema by fixing valid non-sensitive fields and assigning sensitive fields from their allowed domains. Each query is sent to the label-only API while recording the returned label and the end-to-end latency; multiple repeats per query are averaged to reduce noise. Using these (latency, label) pairs, the attacker trains a small timing meta-model $g_{\theta}$ that maps a latency measurement to a probability distribution over labels. The key idea is that input-dependent sparsity and zero-skipping cause some labels to run faster or slower on average, creating a learnable association between latency and label.

\paragraph*{Phase II : Probe the victim by sweeping the sensitive attribute}
For a specific victim with known non-sensitive attributes $\mathbf{x}_{\mathrm{ns}}$, the attacker enumerates the candidate values of the sensitive attribute. For each candidate, the attacker forms the full input by combining $\mathbf{x}_{\mathrm{ns}}$ with that candidate, sends several queries, and records the average latency together with the label returned by the API. This produces a compact per-candidate summary of observed timing and label. This works because changing only the sensitive field can alter activation sparsity and compute, producing measurable timing shifts.

\paragraph*{Phase III : Infer the sensitive attribute and the task label}
For each candidate measured in Phase II, the attacker feeds its latency into $g_{\theta}$ and reads the probability that $g_{\theta}$ assigns to the label returned for that candidate, treating this as a consistency score that reflects how confident that timing is for that label. The candidate with the highest score is reported as the inferred sensitive attribute, and its associated API label is taken as the inferred task label. The intuition is that inference time acts as a proxy for model confidence: the candidate whose timing looks most confident for its label under $g_{\theta}$ is the most plausible.

This scenario directly corresponds to the results in
Table~\ref{tab:leakage_all_summary}, where \scheme{} recovers sensitive
inputs with substantially higher accuracy than
random guessing.

\subsection{Target  B: Inferring the True Output Label}
\label{sec:attack-output}

\paragraph*{Goal}
Infer the true \emph{output} label (e.g., Procedure Code in Texas-100X, Income in Adult) for a specific individual using timing only. Here, the label itself is sensitive (e.g., surgical procedure or diagnosis).

\paragraph*{Assumption}
Similar to Target A, the attacker has no access to confidence scores, the confusion matrix, or marginal priors, and does not know ground-truth labels. The attacker’s input vector for the victim is incomplete: one or more attributes are unknown, so the attacker cannot simply submit the exact record and read off the model’s label.

The adversary follows the same three phases as in Target A to infer the true label using timing only. The difference is that, if the victim record is missing more than one input attribute, Phase II sweeps all feasible combinations of the unknown fields while keeping known attributes fixed, collecting the returned label and averaged latency for each combination. In Phase III, each combination is scored by feeding its latency into $g_{\theta}$ and reading the score assigned to its returned label; the label with the highest score is reported as the inferred true output label.
Intuitively, input-dependent sparsity induces label-specific timing signatures, so the completion whose latency looks most confident for its label under $g_{\theta}$ is most likely to reveal the true label.

\subsection{Target  C: Leakage in Retrieval-Augmented Inference} 
\label{sec:attack-rag}

\paragraph*{Goal}
Infer a sensitive attribute that is never directly present in the
client’s input, but is fetched from a backend or feature store during
inference (RAG-style systems, offline-to-online feature pipelines, or
hospital EHR retrieval).

\paragraph*{Assumption}
Modern AI systems often perform backend data enrichment: given a user ID or partial context, the system retrieves private attributes (e.g., diagnosis codes, risk scores) and feeds them into the model, even though the client never sees these attributes. We assume the attacker can issue query inputs that trigger this retrieval and can measure the model’s timing, but cannot observe the retrieval store itself. The adversary has an auxiliary dataset (disjoint from the training set) for which the sensitive attributes are known. Unlike previous settings, the attacker cannot arbitrarily modify sensitive attributes, because retrieval in the RAG pipeline is fixed by the backend system.


\paragraph*{Step 1 : profiling retrieval-induced timing}
Using auxiliary or synthetic IDs, the attacker repeatedly queries the
system and measures timing. Each ID causes the retrieval subsystem to
fetch a (hidden) set of attributes that feed the model and induce a
distinct sparsity pattern. Over many repetitions, the attacker obtains a
stable timing fingerprint per ID or per attribute value.

\emph{Example.} In a hospital RAG system, different patient IDs
correspond to different hidden procedure histories; timing reveals
whether the model’s decision path includes high-sparsity vs.\ low-sparsity
procedures.

\paragraph*{Step 2 : learning timing clusters for backend attributes}
The attacker clusters timing profiles to discover distinct groups that
correspond to different hidden attribute configurations (e.g., “has high
risk score” vs.\ “low risk score”; “has genetic marker” vs.\ not). If a
small labeled subset of IDs is available (e.g., through leaks or
partial knowledge), the attacker can assign semantic labels to clusters.

\paragraph*{Step 3 : victim inference}
Once timing clusters are established, the attacker observes the timing
of a victim’s query (e.g., the victim’s ID) and assigns it to the
closest timing cluster, thereby inferring properties of the retrieved
backend data. Because retrieval changes the sparsity of the enriched
feature vector, and sparsity directly affects timing, this approach
bypasses output-discretization defenses that assume the only information
channel is the model’s output tokens or probabilities.


Across all three targets, \scheme{} adheres to the same structure:
(1) induce attribute-dependent activations; (2) measure timings to
observe zero-skipping indirectly; (3) learn timing signatures; and
(4) match the victim’s timing to the closest signature. Scenarios A, B,
and C differ in what is considered “sensitive” (input, output, or
retrieved attribute) and what the attacker can control (non-sensitive
features vs.\ keys/IDs). This algorithm is depicted in Figure~\ref{fig:attack-overview}.


\section{Security Analysis}
\label{sec:security-analysis}

This section evaluates \scheme{} under the three scenarios introduced in
Section~\ref{sec:sparsebleed-real}: (A) sensitive input attribute,
(B) sensitive output label, and (C) retrieval-augmented inference
(RAG). For these scenarios, we describe the experimental setup, datasets,
and evaluation metrics, and then present the main findings and how they
tie back to our threat model and root-cause analysis.

\subsection{Experimental Setup}
\label{sec:exp-setup}
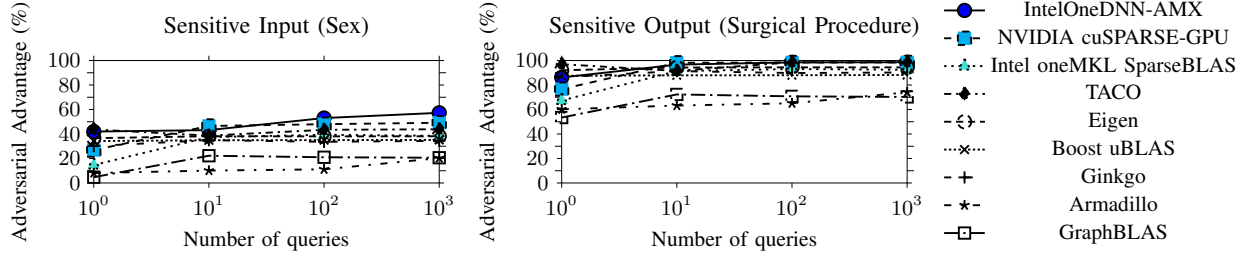
\begin{figure*}[t]
\centering
\scalebox{0.9}{
\begin{tikzpicture}
\begin{groupplot}[
  group style={group size=2 by 1, horizontal sep=1.8cm},
  width=0.28\textwidth, height=1.8cm,
  xmode=log, log basis x=10,
  xmin=1, xmax=1000,
  xtick={1,10,100,1000},
  xlabel={Number of queries},
  ylabel={Adversarial Advantage (\%)},
  ymin=0, ymax=100,
  ytick={0,20,40,60,80,100},
  scale only axis,
  enlarge x limits=false,
  enlarge y limits=false,
  tick align=outside,
  major tick length=2pt,
  minor tick num=1
]

\nextgroupplot[
  title={Sensitive Input (Sex)},
  legend to name=advlegend,
  legend columns=1,
  legend style={draw=none, fill=none, /tikz/every even column/.style={column sep=8pt}}
]

\addplot+[black, solid, mark=*,       thick, mark size=2.8pt, mark options={fill=blue}] coordinates {(1,41.98) (10,43.18) (100,53.02) (1000,57.28)}; \addlegendentry{IntelOneDNN-AMX}
\addplot+[black, dashed, mark=square*,thick, mark size=2.8pt, mark options={fill=cyan}] coordinates {(1,26.88) (10,46.54) (100,48.00) (1000,49.16)}; \addlegendentry{NVIDIA cuSPARSE-GPU}
\addplot+[black, dotted, mark=triangle*,thick, mark size=3.0pt, mark options={fill=Turquoise}] coordinates {(1,14.66) (10,37.44) (100,39.36) (1000,38.50)}; \addlegendentry{Intel oneMKL SparseBLAS}
\addplot+[black, dashdotted, mark=diamond*,thick, mark size=2.8pt, mark options={fill=black}] coordinates {(1,43.72) (10,38.64) (100,43.42) (1000,43.84)}; \addlegendentry{TACO}
\addplot+[black, densely dashed, mark=o, thick, mark size=2.8pt, mark options={fill=white}] coordinates {(1,36.58) (10,38.18) (100,38.04) (1000,38.14)}; \addlegendentry{Eigen}
\addplot+[black, densely dotted, mark=x, thick, mark size=3.0pt] coordinates {(1,34.22) (10,34.90) (100,35.00) (1000,35.18)}; \addlegendentry{Boost uBLAS}
\addplot+[black, loosely dashed, mark=+, thick, mark size=3.0pt] coordinates {(1,30.16) (10,34.98) (100,33.60) (1000,34.36)}; \addlegendentry{Ginkgo}
\addplot+[black, loosely dashdotted, mark=star, thick, mark size=2.6pt] coordinates {(1,8.10) (10,10.12) (100,11.14) (1000,20.32)}; \addlegendentry{Armadillo}
\addplot+[black, dash pattern=on 6pt off 2pt on 1pt off 2pt, mark=square, thick, mark size=2.8pt] coordinates {(1,4.72) (10,22.24) (100,21.00) (1000,20.60)}; \addlegendentry{GraphBLAS}

\nextgroupplot[
  title={Sensitive Output (Surgical Procedure)}
]

\addplot+[black, solid, mark=*,       thick, mark size=2.8pt, mark options={fill=blue}] coordinates {(1,86.16) (10,96.62) (100,98.50) (1000,98.58)}; 
\addplot+[black, dashed, mark=square*,thick, mark size=2.8pt, mark options={fill=cyan}, forget plot] coordinates {(1,76.62) (10,98.14) (100,98.38) (1000,98.42)};
\addplot+[black, dotted, mark=triangle*,thick, mark size=3.0pt, mark options={fill=Turquoise}, forget plot] coordinates {(1,67.36) (10,91.34) (100,93.04) (1000,92.46)};
\addplot+[black, dashdotted, mark=diamond*,thick, mark size=2.8pt, mark options={fill=black}, forget plot] coordinates {(1,97.00) (10,91.48) (100,98.58) (1000,98.62)};
\addplot+[black, densely dashed, mark=o, thick, mark size=2.8pt, mark options={fill=white}, forget plot] coordinates {(1,92.04) (10,94.26) (100,94.40) (1000,94.40)};
\addplot+[black, densely dotted, mark=x, thick, mark size=3.0pt, forget plot] coordinates {(1,87.50) (10,87.90) (100,87.98) (1000,88.18)};
\addplot+[black, loosely dashed, mark=+, thick, mark size=3.0pt, forget plot] coordinates {(1,86.36) (10,91.36) (100,89.78) (1000,90.24)};
\addplot+[black, loosely dashdotted, mark=star, thick, mark size=2.6pt, forget plot] coordinates {(1,60.16) (10,63.22) (100,65.08) (1000,74.04)};
\addplot+[black, dash pattern=on 6pt off 2pt on 1pt off 2pt, mark=square, thick, mark size=2.8pt, forget plot] coordinates {(1,53.42) (10,72.30) (100,70.74) (1000,70.26)};

\end{groupplot}

\node[anchor=west] at ([xshift=0.3cm]group c2r1.east) {\pgfplotslegendfromname{advlegend}};

\end{tikzpicture}
}
\caption{Adversarial Advantage (\%) versus number of queries on the Texas-100X dataset across nine well-known libraries.}
\label{fig:texas_queries_advantage}
\end{figure*}

\noindent\textbf{Hardware and Accelerators.}
We evaluate on three compute backends:
\textbf{(I) NVIDIA A100 data-center GPU};
\textbf{(II) CPU using AVX2/AVX-512 vector units};
\textbf{(III) CPU using Intel AMX tile/matrix units}.
Intel AMX is an integrated on-core accelerator introduced with 4th Gen
Intel Xeon Scalable processors and is specifically designed to speed up
matrix operations in AI and HPC workloads. Our experiments use a server
with Intel Sapphire Rapids microarchitecture, which supports AVX2,
AVX-512, and AMX.

\noindent\textbf{ML Libraries.}
We run the same victim model through a range of dense and sparse
libraries. Table~\ref{tab:libraries-compact} shows the nine stacks we consider.
By instantiating the same logical model across these libraries, we
demonstrate that SparseBleed is not an artifact of a specific software
interface or microarchitectural corner case, but a general phenomenon
that appears whenever sparse-aware linear algebra is used.


\begin{table}[t]
\centering
\footnotesize
\setlength{\tabcolsep}{2pt}
\renewcommand{\arraystretch}{1.05}
\scalebox{0.85}{
\begin{tabular}{l l}
\toprule
\textbf{Library} & \textbf{Backend / Kernel Type} \\
\midrule
oneDNN & CPU (AVX2/512/AMX); Dense GEMM, SpMM \\
oneMKL SparseBLAS & CPU (AVX2/512); SpMV/SpMM (CSR/COO/CSC) \\
cuSPARSE & GPU (A100); SpMV/SpMM/SpGEMM \\
TACO & CPU (AVX2/512); Generated SpMV/SpMM \\
Eigen & CPU (AVX2/512); Dense + CSR SpMV \\
Boost uBLAS & CPU (AVX2/512); Dense/Sparse BLAS ops \\
Ginkgo & GPU/CPU (AVX512); SpMV/SpMM, solvers \\
Armadillo & CPU (AVX2/512); Dense + Sparse wrappers \\
GraphBLAS & CPU (AVX2/512); SpMV/SpMM (semirings) \\
\bottomrule
\end{tabular}
}
\caption{Backends/kernels for evaluated ML libraries.}
\label{tab:libraries-compact}
\end{table}

\noindent\textbf{Victim Model (service).}
For tabular data, we deploy a four-layer feedforward neural network with
two hidden layers of 256 neurons and ReLU activations, following prior
work~\cite{Jayaraman2022Attribute,Shokri2017SP,Jayaraman2021Membership}.
We use ADAM with a learning rate of 0.001 and train one model per
dataset. The inference endpoint is \emph{label-only}: the service
returns the predicted class label but never exposes confidence scores,
logits, or internal activations.
For OrganAMNIST, we train a simple CNN/MLP stack that
classifies 11 organ classes from CT slices.

\noindent\textbf{Datasets.}
We use:

\begin{itemize}[leftmargin=1em]
  \item \textbf{Texas-100X}~\cite{Jayaraman2022Attribute}, a tabular
        hospital dataset with 925{,}128 patient records and 100 surgical
        procedures. For Scenario~A and Scenario~B, we downsample to
        34{,}818 records, selecting 10k for training and 5k for testing.
        The model attains 99.8\% training and 99.1\% test accuracy. For
        Scenario~C, we construct an 81{,}894-record subset covering the
        top-10 diagnoses and top-10 surgery codes.

  \item \textbf{Adult}~\cite{Adult_Dataset}, used in the format of
        Mehnaz et al.~\cite{Mehnaz2022Attribute}. The task is to predict
        income $\le$50k vs.\ $>$50k from 10 socio-demographic and
        work-related attributes. We use 10k training and 5k test
        samples; the model achieves 84.7\% train and 83.7\% test
        accuracy.

  \item \textbf{OrganAMNIST}~\cite{medmnist}, modeling a hospital
        pipeline that retrieves an internal CT slice. The dataset
        contains CT scans from 11 organ classes. We treat the
        fine-grained organ label as a sensitive attribute. We augment
        each sample with synthetic but realistic metadata (age, sex, BMI,
        hospital ID) to simulate a richer clinical query.
\end{itemize}

\noindent\textbf{Preprocessing.}
Categorical features are one-hot encoded; numeric features are
standardized to zero mean and unit variance based on the training split.
The same transform is applied to test and attack inputs, consistent with
prior work~\cite{Jayaraman2022Attribute,Mehnaz2022Attribute}.

\noindent\textbf{Auxiliary Data (Scenario C).}
Unless otherwise stated, the attacker holds approximately 3\% of the
records drawn from the same
population but not used to train the target model. This reflects a
realistic low-knowledge adversary who can observe population-level
statistics but does not have access to the full training dataset or its
sensitive labels.

\noindent\textbf{Metrics.}
We report standard classification metrics: Accuracy, Weighted-Precision,
Weighted-Recall, Weighted-F1, and Adversarial Advantage (normalized gain
over a random guess). Accuracy and Weighted-Recall are identical in our
single-label setting. For timing separation itself, we use Cohen’s $d$
(effect size) to quantify how far apart two timing distributions are;
this is a property of the leakage channel, not of the attack classifier.

\begin{figure}[!htbp]
\centering
\begin{tikzpicture}
\begin{groupplot}[
  group style={group size=2 by 2, horizontal sep=0.65cm, vertical sep=1.2cm},
  width=0.58\columnwidth,
  height=2.8cm,
  ymin=0, ymax=1.0,
  xtick=data,
  xtick pos=left,
  xtick style={draw=none},
  xticklabel style={rotate=25,anchor=east,font=\scriptsize},
  tick label style={font=\scriptsize},
  title style={font=\scriptsize},
  label style={font=\small},
]

\nextgroupplot[
  title={Texas-100X: Input},
  ylabel={Score},
  symbolic x coords={sex,ethnicity,typeadm,race,srcadm},
  xticklabels={sex,ethnicity,type\_adm,race,src\_adm},
  legend to name=mergedlegend,
  legend style={draw=none, fill=none},
  legend columns=4
]
\addplot+[draw=black, thick, mark=square*, mark size=2.3pt, mark options={fill=cyan!70}]
  coordinates {(sex,0.6639)(ethnicity,0.5034)(typeadm,0.2416)(race,0.1510)(srcadm,0.2698)};
\addlegendentry{Acc (W-Rec)}

\addplot+[draw=black, thick, mark=*, mark size=2.3pt, mark options={fill=blue!70}]
  coordinates {(sex,0.7492)(ethnicity,0.6168)(typeadm,0.4838)(race,0.4061)(srcadm,0.5578)};
\addlegendentry{W-Prec}

\addplot+[draw=black, thick, mark=triangle*, mark size=2.5pt, mark options={fill=Turquoise!70}]
  coordinates {(sex,0.6871)(ethnicity,0.5347)(typeadm,0.2824)(race,0.1867)(srcadm,0.3593)};
\addlegendentry{W-F1}

\addplot+[draw=black, dashed, mark=star, mark size=2.2pt, mark options={fill=white}]
  coordinates {(sex,0.5)(ethnicity,0.5)(typeadm,0.2)(race,0.2)(srcadm,0.1111)};
\addlegendentry{Random}

\nextgroupplot[
  title={Texas-100X: Output},
  symbolic x coords={sex,ethnicity,typeadm,race,srcadm},
  xticklabels={sex,ethnicity,type\_adm,race,src\_adm},
]
\addplot+[draw=black, thick, mark=square*, mark size=2.3pt, mark options={fill=cyan!70}, forget plot]
  coordinates {(sex,0.9386)(ethnicity,0.9980)(typeadm,0.9964)(race,0.9976)(srcadm,0.9913)};
\addplot+[draw=black, thick, mark=*, mark size=2.3pt, mark options={fill=blue!70}, forget plot]
  coordinates {(sex,0.9443)(ethnicity,0.9980)(typeadm,0.9964)(race,0.9976)(srcadm,0.9914)};
\addplot+[draw=black, thick, mark=triangle*, mark size=2.5pt, mark options={fill=Turquoise!70}, forget plot]
  coordinates {(sex,0.9388)(ethnicity,0.9980)(typeadm,0.9964)(race,0.9976)(srcadm,0.9913)};
\addplot+[draw=black, dashed, mark=star, mark size=2.2pt, mark options={fill=white}, forget plot]
  coordinates {(sex,0.5)(ethnicity,0.5)(typeadm,0.5)(race,0.5)(srcadm,0.5)};

\nextgroupplot[
  title={Adult: Input},
  ylabel={Score},
  symbolic x coords={mar,edu,race,work,occ},
  xticklabels={marital,education,race,work,occupation},
]
\addplot+[draw=black, thick, mark=square*, mark size=2.3pt, mark options={fill=cyan!70}, forget plot]
  coordinates {(mar,0.5677)(edu,0.4245)(race,0.2399)(work,0.2193)(occ,0.0795)};
\addplot+[draw=black, thick, mark=*, mark size=2.3pt, mark options={fill=blue!70}, forget plot]
  coordinates {(mar,0.5712)(edu,0.4790)(race,0.7481)(work,0.5894)(occ,0.1058)};
\addplot+[draw=black, thick, mark=triangle*, mark size=2.5pt, mark options={fill=Turquoise!70}, forget plot]
  coordinates {(mar,0.5423)(edu,0.4447)(race,0.3507)(work,0.2977)(occ,0.0786)};
\addplot+[draw=black, dashed, mark=star, mark size=2.2pt, mark options={fill=white}, forget plot]
  coordinates {(mar,0.5)(edu,0.3333)(race,0.2)(work,0.1429)(occ,0.0714)};

\nextgroupplot[
  title={Adult: Output},
  symbolic x coords={mar,edu,race,work,occ},
  xticklabels={marital,education,race,work,occupation},
]
\addplot+[draw=black, thick, mark=square*, mark size=2.3pt, mark options={fill=cyan!70}, forget plot]
  coordinates {(mar,0.7986)(edu,0.8274)(race,0.8449)(work,0.8352)(occ,0.8231)};
\addplot+[draw=black, thick, mark=*, mark size=2.3pt, mark options={fill=blue!70}, forget plot]
  coordinates {(mar,0.7858)(edu,0.8194)(race,0.8372)(work,0.8348)(occ,0.8143)};
\addplot+[draw=black, thick, mark=triangle*, mark size=2.5pt, mark options={fill=Turquoise!70}, forget plot]
  coordinates {(mar,0.7892)(edu,0.8215)(race,0.8368)(work,0.8350)(occ,0.8166)};
\addplot+[draw=black, dashed, mark=star, mark size=2.2pt, mark options={fill=white}, forget plot]
  coordinates {(mar,0.5)(edu,0.5)(race,0.5)(work,0.5)(occ,0.5)};

\end{groupplot}

\path (group c1r1.north) -- (group c2r1.north) coordinate[pos=0.5] (LegendPos);
\node[anchor=south, draw=black!60, rounded corners, fill=white, inner sep=1.7pt]
  at ([yshift=4ex]LegendPos) {\scriptsize \pgfplotslegendfromname{mergedlegend}};

\end{tikzpicture}

\caption{Accuracy, weighted precision, recall, and F1 for sensitive 
input (left column) and sensitive output (right column) for Texas-100X (top row) 
and Adult (bottom row). Dashed lines show random-guess baselines.}
\label{fig:metrics_2x2}
\end{figure}
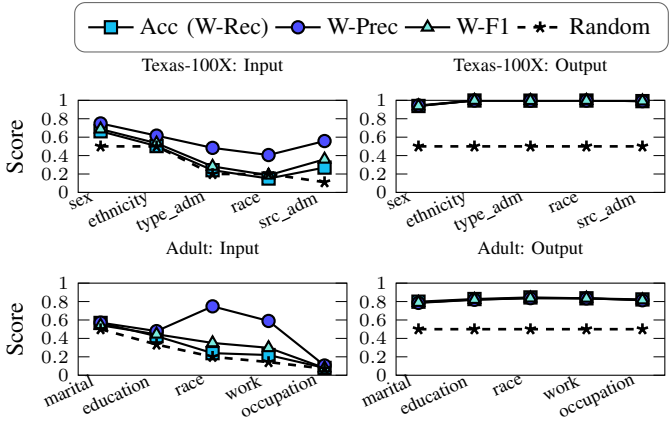

\subsection{Scenario A: Sensitive Input Attribute}
\label{sec:security-input}

In Scenario~A, we apply \scheme{} to infer sensitive \emph{input} fields
from Texas-100X and Adult, such as Sex, Ethnicity, and Race. This
directly instantiates Variant~A from Section~\ref{sec:attack-input}.
For each dataset and attribute:

\begin{enumerate}[leftmargin=1em]
  \item We treat the sensitive field (e.g., Sex) as the target of
        inference and consider the remaining input features as potentially
        known to the attacker.
  \item We query the label-only API for each test record and measure
        timing over $M$ repetitions to form a timing dataset.
  \item Using auxiliary data (synthetic
        samples), we train a lightweight classifier that maps timing
        signatures to
        the target model predicted label (Procedure Code in Texas-100X and Income in Adult).
\end{enumerate}

Table~\ref{tab:leakage_all_summary} (top block) summarizes the leakage
for sensitive inputs. For Texas-100X:
\textbf{Sex (2 classes):} 66.39\% accuracy; W-F1 = 68.71\%; adversarial advantage (AA) = +32.78\%.
\textbf{Source of Admission (9 classes):} 26.98\% accuracy; W-F1 = 35.93\%; AA = +17.85\%.
\textbf{Type of Admission (5 classes):} 24.16\% accuracy; W-F1 = 28.24\%; AA = +5.20\%.
\textbf{Ethnicity (2 classes) and Race (5 classes):} Low accuracy indicates these attributes contribute little to predicting the surgical procedure.

For Adult:
\textbf{Marital Status (2 classes):} 56.77\% accuracy; AA = +13.54\%.
\textbf{Education (3 classes):} 42.45\% accuracy; AA = +13.68\%.
\textbf{Race (5 classes):} 23.99\% accuracy; AA = +4.99\%.
\textbf{Work Class (7 classes):} 21.93\% accuracy; AA = +8.91\%.
\textbf{Occupation (14 classes):} 7.95\% accuracy; AA = +0.87\% (near random).

Taken together with the Texas-100X results, these findings suggest that attributes strongly contributing to prediction of the task (e.g., Sex in Texas-100X and Marital Status in Adult) induce larger and more consistent changes in activation sparsity, which manifest as stronger timing signals. Conversely, attributes that contribute little to prediction of the task (e.g., Race/Ethnicity in Texas-100X) produce smaller, noisier timing effects, yielding near-random inference accuracy.

\begin{table}[t]
\centering
\small
\renewcommand{\arraystretch}{0.9}
\setlength{\tabcolsep}{3pt}  
\scalebox{0.85}{
\begin{tabular}{l l c c c c}
\toprule
\textbf{Dataset} & \textbf{Attr.} & \textbf{Class} &
\textbf{Acc.} & \textbf{W-F1} & \textbf{AA} \\
\midrule
\multicolumn{6}{l}{\textbf{Target A: Sensitive Input Attributes}} \\
Texas-100X & Sex                 &  2 & 66.39 & 68.71 & +32.78 \\
Texas-100X & Ethnicity           &  2 & 50.34 & 53.47 &  +0.68 \\
Texas-100X & Type of Adm.        &  5 & 24.16 & 28.24 &  +5.20 \\
Texas-100X & Race                &  5 & 15.10 & 18.67 &  0.00  \\
Texas-100X & Src. of Adm.        &  9 & 26.98 & 35.93 & +17.85 \\
Adult      & Marital Stat.       &  2 & 56.77 & 54.23 & +13.54 \\
Adult      & Education           &  3 & 42.45 & 44.47 & +13.68 \\
Adult      & Race                &  5 & 23.99 & 35.07 &  +4.99 \\
Adult      & Work Class          &  7 & 21.93 & 29.77 &  +8.91 \\
Adult      & Occupation          & 14 &  7.95 &  7.86 &  +0.87 \\
\midrule
\multicolumn{6}{l}{\textbf{Target B: Sensitive Output Attributes}} \\
Texas-100X & Sex                 &  2 & 93.86 & 93.88 & +87.72 \\
Texas-100X & Ethnicity           &  2 & 99.80 & 99.80 & +99.60 \\
Texas-100X & Type of Adm.        &  5 & 99.64 & 99.64 & +99.28 \\
Texas-100X & Race                &  5 & 99.76 & 99.76 & +99.52 \\
Texas-100X & Src. of Adm.        &  9 & 99.13 & 99.13 & +98.26 \\
Adult      & Marital Stat.       &  2 & 79.86 & 78.92 & +59.72 \\
Adult      & Education           &  3 & 82.74 & 82.15 & +65.48 \\
Adult      & Race                &  5 & 84.49 & 83.68 & +68.98 \\
Adult      & Work Class          &  7 & 83.52 & 83.50 & +67.04 \\
Adult      & Occupation          & 14 & 82.31 & 81.66 & +64.62 \\
\midrule
\multicolumn{6}{l}{\textbf{Target C: RAG Leakage}} \\
OrganAMNIST & CT (Heart)         & 11 & 56.08 & 17.62 & +51.69 \\
OrganAMNIST & CT (Femur-L)       & 11 & 50.37 & 21.48 & +45.41 \\
OrganAMNIST & CT (Femur-R)       & 11 & 48.71 & 20.37 & +43.58 \\
OrganAMNIST & CT (Kidney-L)      & 11 & 45.97 & 43.26 & +40.57 \\
OrganAMNIST & CT (Kidney-R)      & 11 & 43.27 & 41.34 & +37.60 \\
OrganAMNIST & CT (Liver)         & 11 & 43.12 & 43.06 & +37.43 \\
OrganAMNIST & CT (Bladder)       & 11 & 11.60 & 14.71 &  +2.76 \\
Texas-100X  & Surg: C-sec        & 10 & 98.98 & 98.71 & +98.87 \\
Texas-100X  & Surg: Prostate     & 10 & 97.77 & 77.23 & +97.52 \\
Texas-100X  & Surg: Appendix     & 10 & 81.75 & 88.42 & +79.72 \\
Texas-100X  & Surg: Brain        & 10 & 74.27 & 82.21 & +71.41 \\
Texas-100X  & Surg: Gallbladder  & 10 & 74.00 & 65.58 & +71.11 \\
Texas-100X  & Surg: Abdomen      & 10 & 52.23 & 37.56 & +46.92 \\
Texas-100X  & Surg: Derm.        & 10 & 30.48 & 32.16 & +22.76 \\
Texas-100X  & Surg: Blood Ves.   & 10 & 23.83 & 25.19 & +15.37 \\
\bottomrule
\end{tabular}
}
\caption{Inference accuracy (Acc.), weighted F1 (W-F1), 
and adversarial advantage (AA) across all scenarios.}
\label{tab:leakage_all_summary}
\end{table}

Under Scenario~A, \scheme{} achieves non-trivial to high leakage for
several sensitive input attributes across two tabular datasets, without
requiring confidence scores or true labels. The leakage strength is
attribute-dependent, reflecting how strongly each field influences the
activation sparsity we reverse-engineered.

\subsection{Scenario B: Sensitive Output Label}
\label{sec:security-output}

In Scenario~B, the sensitive quantity is the \emph{output} label of the
model, rather than an input field. This aligns with Variant~B from
Section~\ref{sec:attack-output}. We focus on sensitive labels: \textbf{Procedure Code} for Texas-100X and \textbf{Income} for the Adult dataset.
Because the attacker lacks the complete input vector for the victim record, each test withholds one input attribute (as listed in the Attr. column of Table~\ref{tab:leakage_all_summary}) and treats it as unknown to the attacker.
For each dataset:
\begin{enumerate}[leftmargin=1em]
  \item We train a timing-to-label classifier $g(\tau)$ to predict the model’s output from timing alone.
  \item We apply $g(\tau)$ to each sample and take the score corresponding to its predicted label.
  \item To evaluate, we compare inferred labels to ground truth for records that participated in training and report standard metrics (e.g., accuracy, weighted F1, adversarial advantage).
\end{enumerate}

Table~\ref{tab:leakage_all_summary} (middle block) shows that output labels exhibit extremely strong leakage:
On \textbf{Texas-100X}, Sex, Ethnicity, Type of Admission, Source of Admission, and Race are treated as the unknown attributes; under this setting, output labels are recovered with 93.86–99.80\% accuracy.
On \textbf{Adult}, when treating various attributes as unknown (Marital Status, Race, Work Class, and Education), output labels are inferred with $\approx$80–84\% accuracy, and AA ranges from +59.72\% to +68.98\%.

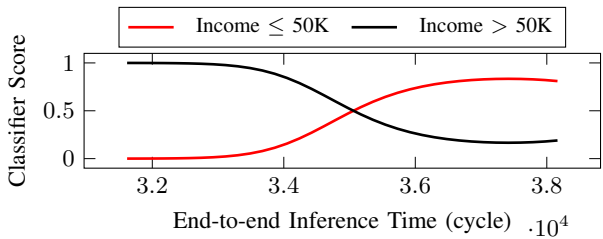
\begin{figure}[!htbp]
\centering

\pgfplotstableread{Diagrams/Data/X_Classifier_Time.txt}\datatableX
\pgfplotstableread{Diagrams/Data/Y_Classifier_Score_1.txt}\datatableYone
\pgfplotstableread{Diagrams/Data/Y_Classifier_Score_2.txt}\datatableYtwo

\pgfplotstablecreatecol[
copy column from table={\datatableYone}{[index]0}
]{yone}\datatableX
\pgfplotstablecreatecol[
copy column from table={\datatableYtwo}{[index]0}
]{ytwo}\datatableX

\begin{tikzpicture}
\begin{axis}[
width=0.95\linewidth,
height=0.35\linewidth,
xlabel={End-to-end Inference Time (cycle)},
ylabel={Classifier Score},
legend style={at={(0.98,1.1)}},
xmajorgrids=false, ymajorgrids=false, xminorgrids=false, yminorgrids=false,
legend columns=2,
legend style={
    at={(0.5,1.05)},
    anchor=south,
    draw=black,
    fill=white,
    font=\footnotesize,                
    column sep=6pt
},
mark=none
]
\addplot[red, line width=1pt] table[x index=0, y=yone] {\datatableX};
\addlegendentry{Income $\leq$ 50K}

\addplot[black, line width=1pt] table[x index=0, y=ytwo] {\datatableX};
\addlegendentry{Income $>$ 50K}
\end{axis}
\end{tikzpicture}

\caption{Timing-to-label classifier score $\mathbb{P}(\hat{y}=c \mid \tau)$ for Intel oneMKL Sparse BLAS timing on the Adult dataset.}
\label{fig:classifier-output}
\end{figure}


Scenario~B demonstrates that sparsity-driven timing leakage is sufficient to recover sensitive labels using only end-to-end timing and the API’s predicted labels. 
Figure~\ref{fig:metrics_2x2} summarizes accuracy, weighted precision/recall, and F1 for inferring sensitive input attributes and sensitive output labels on Texas-100X and Adult; \scheme{} consistently outperforms random guessing. Figure~\ref{fig:classifier-output} shows the timing-to-label classifier’s score computed using the timings from Figure~\ref{fig:two_by_one_sparsity_mkl}.


\subsection{Scenario C: Retrieval-Augmented Inference (RAG)}
\label{sec:security-rag}
Scenario~C evaluates \scheme{} when sensitive attributes are never
explicitly present in the client input, but are fetched from a backend
store (RAG or offline-to-online feature pipelines). This matches
Variant~C from Section~\ref{sec:attack-rag}.

\textbf{OrganAMNIST}: The organ identity is treated as a hidden attribute that is fetched internally (e.g., from a PACS or EHR system).
        The attacker sees only a coarse label or output.
 \textbf{Texas-100X RAG}: The surgical procedure is retrieved
        from a backend table and fed to the model as a hidden feature
        during inference. In both cases, the attacker:

\begin{enumerate}[leftmargin=1em]
  \item Submits a series of queries that trigger retrieval of different
        hidden attributes.
  \item Measures timing $M$ times per sample, forms timing profiles, and
        trains a classifier to map timing to the hidden fine-grained attribute.
\end{enumerate}

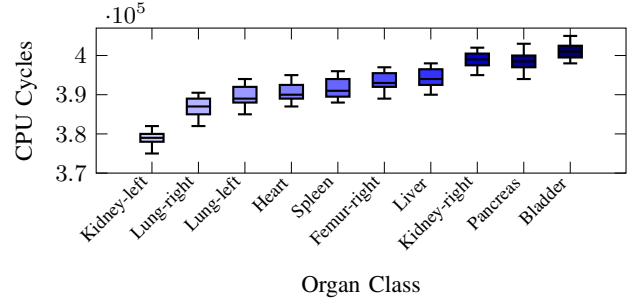
\begin{figure}[t]
\centering
\begin{tikzpicture}

\begin{axis}[
    ymajorgrids=false,
    ylabel={CPU Cycles},
    xlabel={Organ Class},
    width=.97\linewidth,
    height=3.5cm,
    ymin=3.70e5, ymax=4.07e5,
    xtick={1,...,10},
    xticklabels={
        Kidney-left,
        Lung-right,
        Lung-left,
        Heart,
        Spleen,
        Femur-right,
        Liver,
        Kidney-right,
        Pancreas,
        Bladder
    },
    x tick label style={rotate=45, anchor=east, font=\scriptsize},
]

\newcommand{\manualboxplot}[7]{
    \addplot [
        black,
        thick,
        fill=#7
    ] coordinates {
        (#1-0.25,#3)
        (#1-0.25,#5)
        (#1+0.25,#5)
        (#1+0.25,#3)
    } -- cycle;

    \addplot [black, thick] coordinates { (#1-0.25,#4) (#1+0.25,#4) };

    \addplot [black, thick] coordinates { (#1,#3) (#1,#2) };
    \addplot [black, thick] coordinates { (#1,#5) (#1,#6) };

    \addplot [black, thick] coordinates { (#1-0.15,#2) (#1+0.15,#2) };
    \addplot [black, thick] coordinates { (#1-0.15,#6) (#1+0.15,#6) };
}

\manualboxplot{1}{375000}{378000}{379000}{380000}{382000}{blue!20}
\manualboxplot{2}{382000}{385000}{387000}{389000}{390500}{blue!30}
\manualboxplot{3}{385000}{388000}{389000}{392000}{394000}{blue!40}
\manualboxplot{4}{387000}{389000}{390000}{392500}{395000}{blue!50}
\manualboxplot{5}{388000}{389500}{391000}{394000}{396000}{blue!60}
\manualboxplot{6}{389000}{392000}{393000}{395500}{397000}{blue!70}
\manualboxplot{7}{390000}{392500}{394000}{396500}{398000}{blue!80}
\manualboxplot{8}{395000}{397500}{399000}{400500}{402000}{blue!80!black}
\manualboxplot{9}{394000}{397000}{398500}{400000}{403000}{blue!60!black}
\manualboxplot{10}{398000}{399500}{401000}{402500}{405000}{blue!40!black}

\end{axis}
\end{tikzpicture}
\caption{ Inference-time across ten OrganAMNIST classes.  }
\label{fig:organ-timing-boxplots}
\end{figure}

Table~\ref{tab:leakage_all_summary} (bottom block) reports RAG leakage: For \textbf{OrganAMNIST}, leakage is moderate for some organs
        (e.g., Heart at 56.08\%, Femur-Left at 50.37\%) and near random
        (9–11\%) for others (several lung and pancreas classes). This
        indicates that some retrieved organ labels induce more distinctive
        sparsity patterns than others. Figure~\ref{fig:organ-timing-boxplots} shows the timing for OrganAMNIST classes. 
        
        For \textbf{Texas-100X}, we achieve 68.65\% of accuracy, leakage is particularly     strong for
        certain procedures: C-section (98.98\%), Prostate (97.77\%),
        Appendix (81.75\%), Brain (74.27\%), and Gallbladder (74.00\%). Other procedures such as Dermatology and Blood Vessel show weaker yet still non-trivial leakage. In contrast, classes such as Chest and Kidney exhibit near-zero accuracy, largely due to their extremely limited representation in the training distribution.

These findings show that even when sensitive attributes are never exposed in the model’s input or output API, their influence on activation sparsity, and thus on zero-skipping still leaks through timing. Scenario C further demonstrates that \scheme{} extends to RAG and backend-enriched systems: timing alone can partially or fully reveal private fields fetched from backend databases, undermining the common assumption that “only the model output is observable.”

\subsection{Cross-Platform Leakage Across Hardware and Libraries}
\label{sec:security-platforms}

To evaluate whether the leakage depends on a particular hardware stack
or library, we measure adversarial advantage for both sensitive inputs
and outputs across the nine libraries in Table~\ref{tab:libraries-compact}.
Figure~\ref{fig:texas_queries_advantage} shows AA versus the number of queries on Texas-100X for the sensitive input attribute (Sex) and the sensitive output label (Procedure).

Even a single query yields substantial adversarial advantage on many
stacks. For sensitive outputs, the median AA at 1 query is 86.2\%, and
six out of nine libraries exceed 90\% AA by 10 queries, with the best
stacks (TACO, cuSPARSE, IntelOneDNN-AMX) saturating around 98–99\% AA
by 100–1000 queries. For sensitive inputs, AA starts lower but still
reaches 40–60\% on several stacks at modest query budgets.

The leakage is not tied to a single hardware accelerator or library. Both CPUs and GPUs, across multiple dense and sparse libraries, exhibit timing differences large enough to enable practical attacks. 
This reinforces our claim that the root cause lies in sparsity-aware execution itself.




Across all scenarios, datasets, and hardware stacks, the experiments in
this section support the main claims of the paper: (i) activation
sparsity (zero-skipping) creates a strong timing side channel; (ii)
this side channel is rich enough to recover both input and output
sensitive attributes, as well as hidden backend fields; and (iii) the
phenomenon is widespread, spanning multiple models, libraries, and
accelerators.

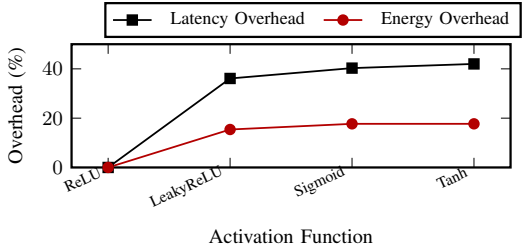
\begin{figure}[!htbp]
\centering
\scalebox{0.85}{
\begin{tikzpicture}
\begin{axis}[
    width=0.95\columnwidth,
    height=3.5cm,
    xlabel={Activation Function},
    ylabel={Overhead (\%)},
    ymin=0, ymax=50,
    symbolic x coords={ReLU,LeakyReLU,Sigmoid,Tanh},
    xtick=data,
    xticklabel style={rotate=25,anchor=east,font=\scriptsize},
    ymajorgrids=false,
    grid style={dotted,gray!40},
    legend columns=2,
    legend style={
        at={(0.55,1.05)},
        anchor=south,
        draw=black,
        fill=white,
        font=\footnotesize,                
        column sep=6pt
    },
    font=\scriptsize,
    line width=1pt,
]

\addplot[
    mark=square*, 
    mark size=2.5pt,
    thick,
    black
] coordinates {
    (ReLU,0)
    (LeakyReLU,36.1)
    (Sigmoid,40.3)
    (Tanh,42.0)
};
\addlegendentry{Latency Overhead}

\addplot[
    mark=*,
    mark size=2.5pt,
    thick,
    red!70!black
] coordinates {
    (ReLU,0)
    (LeakyReLU,15.4)
    (Sigmoid,17.7)
    (Tanh,17.7)
};
\addlegendentry{Energy Overhead}

\end{axis}
\end{tikzpicture}
}
\caption{Overhead of alternative activations compared to ReLU. 
Non-ReLU activations suppress sparsity (and mitigate SparseBleed), 
but incur substantial costs: 36–42\% higher latency and 15–18\% higher energy.}
\label{fig:activation-overhead-line}
\end{figure}

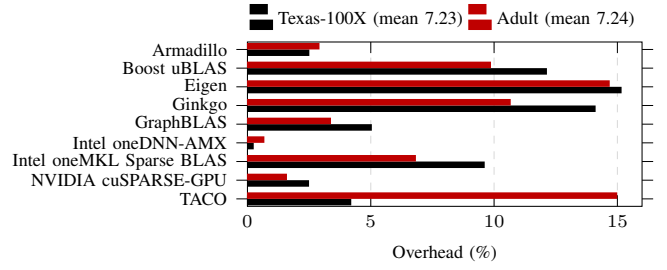
\begin{figure}[t]
\centering
\scalebox{0.98}{
\begin{tikzpicture}
\begin{axis}[
  width=.78\columnwidth,
  height=3.8cm,
  xbar=1pt,
  xmin=0, xmax=16,
  xlabel={Overhead (\%)},
  symbolic y coords={
    Armadillo,
    Boost uBLAS,
    Eigen,
    Ginkgo,
    GraphBLAS,
    Intel oneDNN-AMX,
    Intel oneMKL Sparse BLAS,
    NVIDIA cuSPARSE-GPU,
    TACO
  },
  ytick=data,
  y dir=reverse,
  bar width=1.5pt,
  enlarge y limits=0.05,
  xtick={0,5,10,15},
  xmajorgrids,
  grid style={dashed,gray!30},
  tick label style={font=\scriptsize},
  label style={font=\scriptsize},
  legend style={font=\scriptsize, at={(0.5,1.02)}, anchor=south, draw=none, legend columns=-1}
]

\addplot+[fill=black, draw=black] coordinates
  {(2.4566,Armadillo)
   (12.0906,Boost uBLAS)
   (15.1173,Eigen)
   (14.0641,Ginkgo)
   (4.9908,GraphBLAS)
   (0.20146,Intel oneDNN-AMX)
   (9.5672,Intel oneMKL Sparse BLAS)
   (2.4445,NVIDIA cuSPARSE-GPU)
   (4.1561,TACO)};
\addlegendentry{Texas-100X (mean 7.23)}

\addplot+[fill=red!75!black, draw=red!75!black] coordinates
  {(2.8676,Armadillo)
   (9.8214,Boost uBLAS)
   (14.6366,Eigen)
   (10.6217,Ginkgo)
   (3.3337,GraphBLAS)
   (0.63361,Intel oneDNN-AMX)
   (6.7781,Intel oneMKL Sparse BLAS)
   (1.5494,NVIDIA cuSPARSE-GPU)
   (14.9411,TACO)};
\addlegendentry{Adult (mean 7.24)}

\end{axis}
\end{tikzpicture}
}
\caption{Performance overhead (\%) of the padding defense across libraries for Texas-100X and Adult datasets.}
\label{fig:defense-overhead}
\end{figure}

\section{Mitigation}
\label{sec:mitigation-real}
Output discretization defenses treat inference as an opaque mapping from input to sanitized prediction without taking into account the underlying hardware on which the inference is run. The opacity of inference breaks down in the presence of modern hardware-level sparsity optimizations and activation-gating mechanisms, where large portions of the computation are conditionally skipped based on data-dependent sparsity patterns.
Architectures optimizing sparsity and activation-gating are widely deployed for energy-efficient inference on CPUs, GPUs, and AI accelerators. However, microarchitectural events such as cache-line accesses, power transients, or dataflow scheduling provide a window into the otherwise opaque inference operation before the final output is even produced. Consequently, output discretization defenses fail since they limit only what the model returns, not what the model reveals while computing. Defenses that clip or randomize confidences~\cite{Mehnaz2022Attribute,Jayaraman2022Attribute,AttriGuard2018,Shokri2017SP} remove statistical cues but not timing leakage; \scheme{} remains effective since latency still correlates with sparsity.

\noindent{\textbf{Disabling Zero-Skipping.}} 
Disabling zero-skipping removes the very optimization that yields speed and energy benefits on sparse workloads, and is therefore costly. For example, NVIDIA Sparse Tensor Cores deliver up to 2× math throughput under 2:4 structured sparsity \cite{NVIDIAA100}; turning this off can impose up to a 100\% latency increase on the affected kernels. Beyond performance, energy also rises because more MACs and memory transactions are executed. In our AMX microbenchmarks (hardware-only, without any library optimizations), we observed that when zero-skipping cannot be leveraged, power increases relative to the sparsity-optimized case, yielding an overhead of up to 25\%.


\noindent{\textbf{Software-Level Defense.}}
We consider a padding defense in which the API withholds the response until a fixed budget equal to the worst-case end-to-end inference time elapses. This masks timing variations without performing extra computation; the additional energy is largely limited to idle power during the padding interval. Figure~\ref{fig:defense-overhead} reports performance overhead across nine libraries on Texas-100X (80–90\% sparsity) and Adult (40–60\% sparsity): the average performance overhead is 7.24\%. While effective against remote timing adversaries, a co-located privileged attacker who can observe shared resources (e.g., contention or performance counters) could still estimate true runtime; stronger isolation or additional countermeasures are required in that threat model.


\noindent{\textbf{Different Activation Functions.}} We can mitigate sparsity-based leakage by replacing ReLU with alternative activation functions such as LeakyReLU, Sigmoid, or Tanh. However, Figure~\ref{fig:activation-overhead-line} shows that eliminating ReLU’s zero-inducing behavior imposes a substantial efficiency cost: performance drops by 36.1\%, 40.2\%, and 42.1\%, and power efficiency decreases by 15.4\%, 17.7\%, and 17.7\% for LeakyReLU, Sigmoid, and Tanh, respectively.


\noindent{\textbf{Leakage-Aware Training.}}  The training procedure can be modified by changing the loss function to include an error term related to the demonstrated leakage.  The training will then attempt to co-optimize for both accuracy and side-channel resilience.  The idea of changing training has been shown in the context of reducing power-based physical side-channel leakage~\cite{dubey2025stochastictrainingsidechannelresilient} and reverse-engineering resilience of neural networks~\cite{kurian2025traindefenddefensecryptanalytic}, but needs to be adapted for sparsity-driven architectural vulnerabilities that we discussed in this work.

\section{Conclusion}
\label{sec:conclusion-real}
Trained models exhibit attribute-dependent activation sparsity, and sparse-aware stacks translate these differences into measurable end-to-end timing via zero-skipping. Leveraging this signal, \scheme{} infers sensitive input attributes and true output labels in label-only settings and extends to retrieval-based pipelines. Disabling sparsity acceleration is costly, whereas software-level padding to the worst-case time masks timing with modest average overhead and near-zero extra energy. In parallel, sparsity-aware training that reduces attribute-driven sparsity gaps can further diminish leakage.

\bibliographystyle{IEEEtranS}
\bibliography{refs}

\end{document}